\documentclass[11pt,a4paper]{article}

\usepackage{xcolor}
\usepackage[left= 2cm, right= 2cm]{geometry}
\usepackage{amsmath,latexsym,amssymb,slashed}
\usepackage{graphicx}				
\usepackage{amssymb}					
\usepackage{tabularx}				
\usepackage[vcentermath]{youngtab}	

\usepackage[latin1]{inputenc}
\usepackage[T1]{fontenc}

\newcommand{\eq}[1]{\begin{equation}#1\end{equation}}
\newcommand{\spl}[1]{\begin{split}#1\end{split}}

\def\bea{\begin{eqnarray}}
\def\eea{\end{eqnarray}}

\def\nn{\nonumber}

\def\d{\text{d}}

\newcommand{\cp}[1]{\mathbb{CP}^{#1}}

\begin{document}
\setlength{\parindent}{0pt}

\begin{titlepage}
	\vspace{14pt}
	
	\begin{center}
		
		{\Large \bf D-branes  and non-Abelian T-duality}\\

		\vspace{1.6cm}
		
		{\bf \footnotesize Robin Terrisse $^{a,}$\footnote{rterrisse@ipnl.in2p3.fr, $^2$tsimpis@ipnl.in2p3.fr, $^3$cwhiting@bates.edu}, Dimitrios Tsimpis${^{a,2}}$, and Catherine A. Whiting$^{b,c,3}$ }
		
		\vspace{1cm}
		
		{\bf }

		\vspace{.2cm}
		
		{\it ${}^a$  Institut de Physique Nucl\'eaire de Lyon  }\\
		{\it Universit\'e de Lyon, UCBL, UMR 5822, CNRS/IN2P3 }\\
		{\it 4 rue Enrico Fermi, 69622 Villeurbanne Cedex, France  }\\
		
		\vspace{.4cm}
		
		{\it ${}^b$ National Institute for Theoretical Physics}\\
		{\it School of Physics and Mandelstam Institute for Theoretical Physics}\\
		{\it University of the Witwatersrand, Johannesburg}\\
		{\it WITS 2050, South Africa}
		
		\vspace{.2cm}
		
		{\it ${}^c$ Department of Physics and Astronomy, Bates College}\\
		{\it Lewiston, ME 04240, USA}
		
		\vspace{14pt}

	\end{center}
	\begin{abstract}
	\noindent
We study the effect of non-Abelian T-duality (NATD) on D-brane solutions of type II supergravity. Knowledge of the full  brane solution allows us to track the brane charges and the corresponding brane configurations, thus providing justification for brane setups previously proposed in the literature and for the common lore that Dp brane solutions give rise to D(p+1)-D(p+3)-NS5 backgrounds under SU(2) NATD transverse to the brane. 
In brane solutions where spacetime is empty and flat at spatial infinity before  NATD, the spatial infinity of the NATD is universal, i.e.~independent of the initial brane configuration. Furthermore, it gives enough information to determine the ranges of all coordinates after NATD.  
In the more complicated examples of the D2 branes considered here, where spacetime is not asymptotically flat  before NATD,  
the interpretation of the dual solutions remains unclear. 
In the case of supersymmetric D2 branes arising from M2 reductions to IIA on Sasaki-Einstein seven-manifolds, we explicitly verify that the solution
obeys the appropriate generalized spinor equations for a supersymmetric domain wall in four dimensions. 
We also investigate the existence of supersymmetric mass-deformed D2 brane solutions.
	\end{abstract}

\end{titlepage}


\setcounter{footnote}{0}

\tableofcontents


\newpage

\section{Introduction}

Non-Abelian T-duality (NATD) has recently attracted renewed interest, not least as a tool for generating new supergravity solutions. A partial list of related literature includes \cite{Ossa:1992vc,Fridling:1983ha,Fradkin:1984ai,Sfetsos:2010uq, Lozano:2011kb,Itsios:2013wd, Lozano:2012au,Lozano:2014ata, Bea:2015fja, Lozano:2015bra,Macpherson:2014eza,Zayas:2015azn,PandoZayas:2017ier}. 
One problem in this context is that NATD only carries local information: even when the starting point (the ``seed'') 
is a globally well-defined solution, NATD will typically generate a highly complicated local solution whose global completion, if it exists, is completely obscured. 
The main motivation of the present paper is to consider, as seeds of the NATD, full-fledged brane solutions --as opposed to AdS near-horizon solutions, which has been the case before. 
The point is that, as we will see, being able to follow the interpolation between the near-horizon and spatial infinity limit, gives us a better handle on the brane configuration and the global properties of the dual.

Before NATD, standard intersecting brane solutions (i.e.~those following the simple harmonic superposition rule) often interpolate between two asymptotic regions, 
each of which is an independent supergravity solution in its own right:
 flat space at spatial infinity, and the near-horizon limit --which in the examples considered here always contains an AdS factor. In fact the brane is not strictly-speaking present in these solutions: the brane backreacts on the flat space in which it is initially inserted and dissolves into flux, so that the resulting solution is without sources.\footnote{
 Nevertheless the harmonic superposition rule allows us to trace the original source (i.e.~before backreaction) as a delta function defined in flat space.} 
In the interpolating and near-horizon solutions there is still a remnant of the brane, whose charge can be computed by integrating the flux.

\begin{figure}[tb!]
\begin{center}
\includegraphics[width=0.65\linewidth]{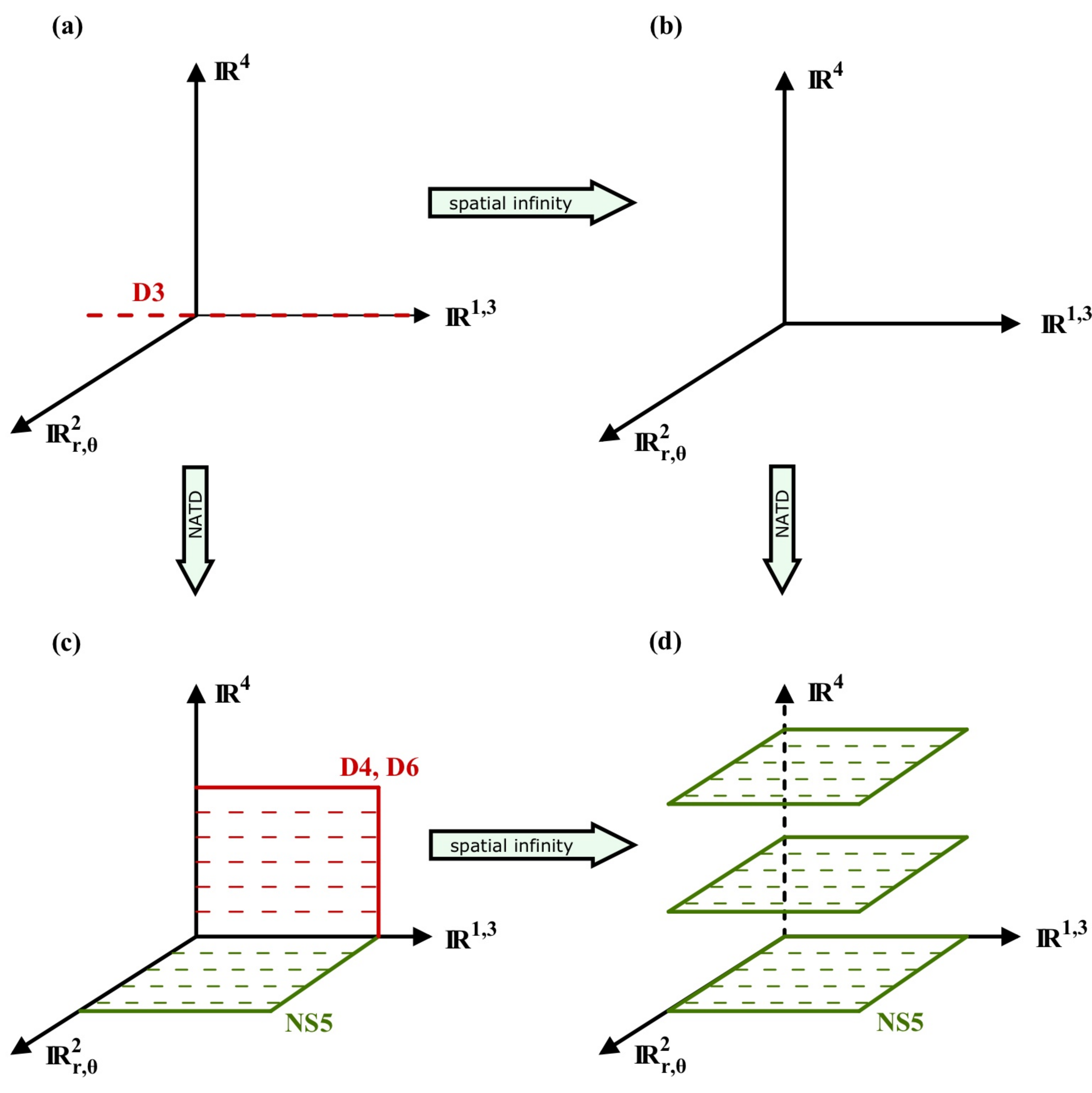}
\end{center}
\caption{\small (a) 
D3 brane in flat space; (b) empty flat space; (c) configuration of intersecting D4-D6-NS5 branes; (d) continuous linear distribution of  NS5 branes along a half line. 
The spatial infinity limit of the dual solution is defined by taking the limit before the NATD.} \label{fig1}
\end{figure}

The NATD of the spatial infinity limit of a standard intersecting brane solution is then universal, i.e. it is the same for all standard intersecting brane solutions: it is simply the NATD of flat space. As we will see in detail in section \ref{sec:24}, in this case an $SU(2)$ NATD  generates a distribution, whose density can be determined explicitly, of parallel NS5 branes continuously distributed along a half line. However after NATD the notion of spatial infinity limit and near-horizon limit are no longer necessarily meaningful. Still, as previously stated, both limits are genuine solutions so that NATD will generate new solutions out of them. These will be called respectively the spatial infinity and near-horizon limit of the dual. For the case of the D3 brane,  discussed in section \ref{sec:2}, this definition leads to the commutative diagram of figure \ref{fig1}.

Furthermore, all examples considered here will be seen to be consistent with  the common lore that Dp brane solutions give rise to D(p+1)-D(p+3)-NS5 backgrounds under SU(2) NATD transverse to the brane \cite{Lozano:2016wrs}
\footnote{Some early examples where this has been hinted at through specific examples are in \cite{Lozano:2014ata,Lozano:2015bra}, but it was proposed generally in \cite{Lozano:2016wrs}.}.

More generally, for nonstandard brane solutions, such as the D2 branes considered here in  sections \ref{sec:5}, \ref{sec:4}, \ref{sec:3}, the geometry may not asymptote to flat space even before NATD. 
In that case the original solution and its dual will in general contain non-vanishing fluxes even at spatial infinity. 
Nevertheless, by zooming in near the locus of the NS5, one can still see the presence of a continuous distribution of NS5 branes in accordance with the harmonic superposition prescription for intersecting branes.

Having the full interpolating solution can facilitate reading off the possible global completions of the geometry after NATD. In particular the topology of the slices $r=\text{constant}$  can be studied more easily by taking the spatial infinity limit $r\rightarrow\infty$, where the various expressions simplify. As already mentioned, this becomes most clear in the standard cases where, before the NATD, the 
space becomes flat at spatial infinity and the NATD is universal.

In order to describe the supersymmetric D2-branes of our paper arising from M2 reductions to IIA on Sasaki-Einstein seven-manifolds, we used the formalism of generalized geometry and the seven-dimensional pure spinors of \cite{Haack:2009jg}. We explicitly verify that this class of solutions
obeys   the appropriate generalized spinor equations for a supersymmetric domain wall in four dimensions. 
One of our main motivations in deriving these supersymmetric domain wall equations was to search for the brane solution, if it exists, whose near horizon limit corresponds to the massive IIA solution found in \cite{Lust:2009mb}.  The near horizon limit of the D2 brane solution we study in section \ref{sec:5}  corresponds to the massless limit and provides a simple solution to these supersymmetry equations in the case of SU(3) structure.
A longer term goal is to understand whether the massive IIA deformation of the backgrounds in section \ref{sec:5} are related to a class of massive IIA $AdS_4$ solutions \cite{Rota:2015aoa} which arise as a compactification from $AdS_7$, and are dual to 3d twisted compactifications of 6d (1,0) SCFTs.  This work was initiated  as a first step towards perhaps a better understanding of the mass deformation solution in \cite{Lust:2009mb} and its NATD found in  \cite{PandoZayas:2017ier} in that context. 
While we were unable to find that interpolating mass deformed solution in this paper, by investigating the existence of supersymmetric mass-deformed D2 brane solutions we have ruled out a large class of Ans\"{a}tze.

The outline of the  paper is as follows. We revisit the case of the D3 brane in section \ref{sec:2}. Various D2 brane solutions are examined in sections \ref{sec:5}, \ref{sec:4}, \ref{sec:3}. 
These are obtained by reduction from eleven dimensional solutions of M2 branes transverse to cones over $S^7$ or $Y^{p,q}$ reviewed in section \ref{sec:0}.  
In section \ref{sec:dw} we examine  massive deformations of the 
supersymmetric massless IIA D2 brane solutions of section \ref{sec:5}. We start by casting the supersymmetry equations of the branes in the formalism of generalized geometry 
in section \ref{sec:susydw}. Massive deformations of the resulting pure spinor equations are examined in section \ref{sec:susymd}. We conclude in section \ref{sec:concl}. In the appendix we explain our various conventions and compare them to the literature.


\section{D3 brane}\label{sec:2}


The metric describing a stack of parallel D3 branes is given by,
\bea \label{D3Metric}
\d s^2=H(r)^{-1/2}\d s^2(\mathbb{R}^{1,3})+H(r)^{1/2}[\d r^2+r^2\d s^2(S^5)]~,
\eea
where $H(r)=(1+\frac{L^4}{r^4})$. The $S^5$ is parameterized as follows,
\bea \label{S5la}
ds^2(S^5)=\d\alpha^2+\sin^2\!\alpha ~\!\d\theta^2 +\tfrac{1}{4}\cos^2\!\alpha~\!(\sigma_1^2+\sigma_2^2+\sigma_3^2)
~,
\eea
where $\alpha\in[0,\tfrac{\pi}{2}]$, $\theta\in[0,2\pi]$, 
and $\sigma_i$  are left-invariant $SU(2)$ Maurer Cartan one-forms given by,
\bea\label{si}
\sigma_1&=&-\sin\psi_1\d\theta_1+\cos\psi_1\sin\theta_1\d\phi_1\nn\\
\sigma_2&=& \cos\psi_1 \d\theta_1+\sin\psi_1\sin\theta_1\d\phi_1 \nn \\
\sigma_3&=&\cos\theta_1 \d\phi_1+\d\psi_1~,
\eea
with ranges $\psi_1\in[0,4\pi]$, $\theta_1\in[0,\pi]$, $\phi_1\in[0,2\pi]$. 
This background is supported by a constant dilaton and an $F_5$ flux given by,
\bea \label{D3F5}
F_5=(1+\star)\d x_0\wedge \d x_1\wedge \d x_2\wedge \d x_3\wedge \d H(r)^{-1}=\d x_0\wedge \d x_1\wedge \d x_2\wedge \d x_3\wedge \d H(r)^{-1}-4 L^4\d\Omega_5~.
\eea
Upon quantization of the five-form flux, one obtains the well-known relation between the constant $L$ in the harmonic function and the number $N_{D3}$ of D3 branes: $L^4=4\pi\alpha'^2N_{D3}$.

The D3 branes lie along the $\mathbb{R}^{1,3}$ directions. This can be seen in a probe approach. Consider the same expression as (\ref{D3F5}) for a flux living now in $\mathbb{R}^{1,9}$. The coordinates now refer to the metric (\ref{D3Metric}), but with $H=1$. Since in spherical coordinates the $S^5$ collapses at $r=0$, its volume form $\d\Omega_5$ is ill-defined. However $F_5$ is a well-defined current (i.e. a distribution-valued form) and we can compute:
\eq{\label{D3trans}\d F_5 = \d\star F_5 = 4L^4 \delta(r) \d r\wedge \d\Omega_5}
%
%

This means that a brane is inserted in $r=0$. In this coordinate system this is a codimension 6 space, and thus a D3 lying along $\mathbb{R}^{1,3}$. In the transverse space $\mathbb{R}^6$ the brane looks like a point.

The D3 now acts as a source for the flux $F_5$, which backreacts on the metric through Einstein's equations to give (\ref{D3Metric}). The global geometry has changed, and the $S^5$ no longer collapses. The supergravity equations are solved without sources and the brane cannot be seen anymore. Nevertheless the information about the brane is still present in the charge carried by the flux.

\subsection{Near-Horizon and spatial infinity}

Taking (\ref{D3Metric}) as an ansatz for the metric, the supergravity equations reduce to an equation on $H$. In the probe interpretation this amounts to saying that $H$ is harmonic in the transverse space. If we further constrain $H$ to depend on $r$ only, the general solution is of the form,
\eq{H(r)=  a+\frac{b}{r^4} 
~,}
where $a$ and $b$ are two integration constants. $b$ can readily be interpreted as the brane charge. Then two limiting cases arise:

\paragraph{Spatial infinity:} If $b=0$, $H$ is a constant and the space is flat without flux: no brane is inserted. Since $H \rightarrow a$ when $r\rightarrow \infty$, this case is called the spatial infinity limit.

\paragraph{Near Horizon:} If  $a=0$, the solution becomes,
\bea
\d s^2&=&\d s^2(AdS_5)+L^2\d s^2(S^5)=\frac{r^2}{L^2}\d s^2(\mathbb{R}^{1,3})+\frac{L^2}{r^2}\d r^2+L^2\d s^2(S^5)\nn \\
F_5&=&(1+\star)\frac{4r^3}{L^4}\d x_0\wedge \d x_1\wedge \d x_2\wedge \d x_3\wedge \d r~,
\eea
which is the well-known $AdS_5\times S^5$ background. For $r\rightarrow 0$, $H\sim \frac{b}{r^4}$ so that this case is called the near horizon limit.

It is remarkable at first sight that both limits correspond to genuine backgrounds. The reason behind this is that they ultimately correspond to different choices of integration constants. These considerations might seem trivial for now, but they will be relevant in the following, when the brane configurations become more involved.

%

\subsection{The NATD}\label{sec:21}

After performing NATD along the SU(2) isometry in the $\sigma_i$, cf.~\eqref{si}, the background \eqref{D3Metric}, \eqref{D3F5} becomes,
\bea\label{D3NATD}
{}{ds}^2&=&H(r)^{-1/2}ds^2(\mathbb{R}^{1,3})+H(r)^{1/2}[dr^2+r^2(d\alpha^2+\sin^2\alpha d\theta^2)]\nn \\&~&+\frac{1}{4}\Big[\frac{\alpha'^2}{\Xi}d\rho^2+\frac{\Xi^2}{64\alpha'\Delta}\rho^2(d\chi^2+\sin^2\chi d\xi^2)\Big]\nn \\
{}{B}_2&=&-\frac{\rho^3\Xi}{256\Delta}\sin\chi d\chi\wedge d\xi \nn \\
e^{-2{}{\phi}}&=&\Delta, \quad \Delta =  \frac{\Xi}{64\alpha'^3}(\alpha'^2\rho^2+\Xi^2),\quad \Xi=r^2\cos^2\alpha\sqrt{H(r)}~,
\eea
and nonzero RR fluxes given by,
\bea\label{D3NATDRR}
{}{F}_2&=&-\frac{\Xi}{8\alpha'^{3/2}}\frac{H'(r)}{\sqrt{H(r)}}r^3\cos\alpha\sin\alpha d\alpha\wedge d\theta\nn \\
{}{F}_4&=&\frac{\Xi^2}{2048\alpha'^{3/2}\Delta}\frac{H'(r)}{\sqrt{H(r)}}r^3\rho^3\cos\alpha\sin\alpha\sin\chi d\alpha\wedge d\theta\wedge d\chi\wedge d\xi~.
\eea
In particular we see that the NATD has resulted in a metric which is singular at $\alpha=\tfrac{\pi}{2}$. Moreover the duality has generated a nonvanishing Kalb-Ramond field $B_2$ and a varying dilaton $\phi$.

Note that the background (\ref{D3NATD}) contains a family of solutions, inheriting its degrees of freedom from the $D3$ solutions before duality: for each choice of harmonic function $H$, NATD generates a different solution. We will keep the same denomination for the different limits, namely the near-horizon for $H=\frac{L^4}{r^4}$ and spatial infinity for $H=1$ (i.e. $L=0$). However their interpretation as different limits of the interpolating dual background is less meaningful. We will study them separately to get a better view on the brane configurations.

For later use let us rewrite  the metric in \eqref{D3NATD} in terms of the coordinates defined by,
\eq{\label{allcoord}
 x=r\sin\alpha\cos\theta~;~~~
  y=r\sin\alpha\sin\theta~;~~~
  u=r\cos\alpha~.
}
Recalling the ranges of the $\alpha$, $\theta$ coordinates, cf.~\eqref{S5la}, 
 we see  that $u\geq0$, while $x,y\in\mathbb{R}$.   
Simplifying with $r^2=x^2+y^2+u^2$ then gives,
\eq{\label{blabla}
ds^2=H^{-1/2}\Big[ds^2(\mathbb{R}^{1,3})+\frac{\alpha'^2}{4u^2}d\rho^2\Big]
+H^{1/2}\Big[dx^2+dy^2+du^2+\frac{\alpha'^2\rho^2 u^2}{4(\alpha'^2\rho^2+H u^4)}(d\chi^2+\sin^2\chi d\xi^2) \Big]
~.}
In these coordinates, the metric is singular at $u=0$.

\subsubsection{Brane configuration and charges}\label{sec:235}

The non-vanishing fluxes might indicate the presence of branes. Here we could expect NS5, D4 and D6 branes as magnetic sources for $H$, $F_4$ and $F_2$. The first clue is given by the corresponding charges.

Let us start with the NS flux. An appropriate cycle would be the following: start at constant $\alpha=\alpha_0$ and integrate along $\rho,\chi,\xi$ where $\rho$ goes from $0$ to $\rho_0$. At $\rho=0$ the cycle closes but we need to close it at $\rho_0$. To do so, keep $\rho$ constant and vary $\alpha$ from $\alpha_0$ to $\pi/2$. The resulting charge will be independent of $\alpha_0$ so we can take the limit $\alpha_0\rightarrow \pi/2$.

Along the cycle $\left(\Sigma_3=[\rho,\chi,\xi], \alpha=\frac{\pi}{2}\right)$, $H_3$ simplifies to 
\bea \label{B2H3D3}
H_3&=&\frac{1}{4}\alpha' \sin\chi d\xi\wedge d\chi\wedge d\rho~.
\eea
Integrating $H_3$ yields,
\bea\label{QNS5D3}
Q_{NS5}=\frac{1}{2\kappa_{10}^2T_{NS5}}\frac{\alpha'}{4}\int_0^{\rho_0}d\rho\int_0^{\pi}\sin\chi d\chi\int_0^{2\pi}d\xi=\frac{\rho_0}{4\pi}=N_{NS5}~,
\eea
In fact the charge will depend only on the value of $\rho$ when the cycle reaches $\alpha=\pi/2$. As we will see more explicitly in the simpler case of the spatial infinity limit in section \ref{sec:24}, 
this suggests a continuous distribution of NS5 branes at $\alpha=\pi/2$ along the $\rho$ direction, with constant charge density. For the flux to be quantized we need to close the cycle at quantized values of $\rho$, namely $\rho_0=4n \pi$.
The NS5 branes are thus located at the singularity: this can be seen from the form of the metric and NS-NS fields in the limit $\alpha\rightarrow \frac{\pi}{2}$, which is consistent 
with the general form expected from the harmonic superposition rule \cite{Youm:1997hw}.  After defining $\nu=(\pi/2-\alpha)^2$ we find, in the $\alpha\rightarrow\frac{\pi}{2}$ limit,
\eq{\spl{\label{NS5metric}
ds^2&=H^{-1/2}ds^2(\mathbb{R}^{1,3})+H^{1/2}\Big(dr^2+r^2d\theta^2+\frac{r^2}{4\nu}\Big[ d\nu^2 +\frac{\alpha'^2}{Hr^4}d\rho^2+\nu^2(d\chi^2+\sin^2\chi d\xi^2)\Big]\Big)\\
e^{2{\phi}}&=  \frac{64\alpha'}{r^2\sqrt{H}\rho^2\nu}~;~~~H_3= \frac{\alpha'}{4}\sin\chi d\rho\wedge d\chi\wedge d\xi~.
}}
The harmonic function in the space transverse to the NS5 is proportional to $\nu^{-1}$, indicating the presence of NS5 branes at $\nu =0$. However this is not a point in the transverse space.  Since $\rho$ is still unconstrained, this is consistent with a distribution of charge along $\rho$.

In order to determine the configuration of the remaining branes we follow the same strategy.
%
%
Recall that in solutions with nonzero $B_2$, the quantized charges are the Page charges, defined as integrals of the Page forms,
\bea
\tilde{F}_p=F_pe^{-B_2}.
\eea
As can be seen from this definition, the  Page charges depend on the cohomology class of $B_2$, i.e.~they are not invariant under large gauge transformations of $B_2$.

Integrating the Page forms in the D3 brane solution gives,
\bea\label{QD6D3}
Q_{D6}&=&\frac{1}{2\kappa_{10}^2T_{D6}}\frac{L^4}{2\alpha'^{3/2}}\int_0^{\frac{\pi}{2}}\cos^3\alpha\sin\alpha d\alpha \int_0^{2\pi}d\theta=N_{D6}\\
Q_{D4}&=&0~,
\eea
which leads to $L^4=8\alpha'^2N_{D6}$. If we denote by  $\Delta Q_{D4}$ the change of D4 brane charge under a large gauge transformation of $B_2$,
\bea\label{deltaB}
\Delta B_2=-n\pi\alpha'\sin\chi d\xi\wedge d\chi~,
\eea
we find,
\bea
\Delta Q_{D4}&=&\frac{1}{2\kappa_{10}^2T_{D4}}\int -\Delta B_2\wedge F_2 \nn \\&=&\frac{1}{2\kappa_{10}^2T_{D4}}\frac{n\pi L^4}{8\sqrt{\alpha'}}\int_0^{\frac{\pi}{2}}\cos^3\alpha\sin\alpha d\alpha\int_0^{\pi}d\theta\int_0^{\pi}\sin\chi d\chi\int_0^{2\pi}d\xi\nn \\&=&\Delta N_{D4}~,
\eea
which leads to $L^4=\frac{1}{n}8\alpha'^2\Delta N_{D4}$.  From this we readily see that 
\bea \label{deltaQD4}
\Delta Q_{D4}=nN_{D6}.
\eea 
This is nothing other than the creation of D4 branes via a Hanany-Witten effect \cite{Hanany:1996ie}, as will be reviewed in the following in section \ref{sec:23}.
In order to get a probe interpretation of these brane charges we would need to know in which background the branes are inserted, but the situation is not entirely clear here. The expression for the fluxes suggests that the $D6$ is transverse to $r,\alpha,\theta$ and that the $D4$ is transverse to $r,\alpha,\theta,\chi,\xi$. This would lead to the following brane configuration:
\begin{center}
	\begin{tabular}{ |c|cccc|cccccc|} 
		\hline
		& 0&1&2&3 &$ r$&$\alpha$&$\theta$&$\rho$&$\chi$&$\xi$ \\
		\hline
		$NS5$ & $\times$ &$\times$ &$\times$ &$\times$ & $\times$ & &$\times$ & & & \\ 
		$D6$  & $\times$&$\times$&$\times$&$\times$ & & & &$\times$ &$\times$ & $\times$\\ 
		$D4$ & $\times$&$\times$& $\times$ &$\times$ &  & & &$\times$ & & \\ 
		\hline
	\end{tabular}
\end{center}

\subsubsection{Spatial infinity limit}\label{sec:24}

The spatial infinity limit gives the following background:
\bea\label{cxvxcv}
{ds}^2&=&ds^2(\mathbb{R}^{1,3})+dr^2+r^2(d\alpha^2+\sin^2\alpha d\theta^2)+\frac{1}{4}\Big[\frac{\alpha'^2}{\Xi}d\rho^2+\frac{\Xi^2}{64\alpha'\Delta}\rho^2(d\chi^2+\sin^2\chi d\xi^2)\Big]\nn \\
{B}_2&=&-\frac{\rho^3\Xi}{256\Delta}\sin\chi d\chi\wedge d\xi \nn \\
e^{-2{\phi}}&=&\Delta, \quad \Delta =  \frac{\Xi}{64\alpha'^3}(\alpha'^2\rho^2+\Xi^2 ),\quad \Xi=r^2\cos^2\alpha~.
\eea
Here there are no RR fluxes anymore so all the D-brane charges vanish. The configuration is thus much simpler. In fact it will now be possible to understand the exact brane configuration, as is done for the D3. Moreover this background is the NATD of the spatial infinity limit of the D3 brane solution: the NATD (\ref{cxvxcv}) is  simply the NATD of flat space along an $S^3\subset\mathbb{R}^4$ factor. 
This decomposition is thus better suited for the spatial infinity limit than the $S^5\subset\mathbb{R}^6$ decomposition of the D3 brane solution. Accordingly  
the seed metric before NATD reads,
\bea
\d s^2=\d s^2(\mathbb{R}^{1,5})+\d u^2+u^2\d s^2(S^3)~,
\eea
which is simply the spatial infinity limit of the metric \eqref{D3Metric}  written in the coordinates of \eqref{allcoord}. 
In these coordinates the  NATD metric \eqref{cxvxcv} is given by,
\bea
ds^2&=&ds^2(\mathbb{R}^{1,5})+ du^2 + \frac{\alpha'^2}{4u^2}d\rho^2+ \frac{\alpha'^2\rho^2 u^2}{4(\alpha'^2\rho^2+ u^4)}(d\chi^2+\sin^2\chi d\xi^2)
~.
\eea
Let us now make a further change of variable,
\bea
u &=& R^{1/4} \sqrt{\sin \frac{\theta}{2}} \\
\alpha' \rho &=& R^{1/2} \cos\frac{\theta}{2}~,\nn
\eea
upon which  the metric becomes,
\bea\label{clearly}
ds^2&=&ds^2(\mathbb{R}^{1,5})+ \frac{1}{16R^{3/2}\sin\frac{\theta}{2}}\left[dR^2 + R^2 d\theta^2+ R^2\sin^2\theta(d\chi^2+\sin^2\chi d\xi^2)\right] \\
    &=& ds^2(\mathbb{R}^{1,5}) + f(R,\theta) ds^2(\mathbb{R}^{4})~,
\eea
where in order to obtain a complete metric on $\mathbb{R}^4$ we must have $\theta\in[0,\pi]$. 
In the second line above  we have introduced the function,
\eq{ \label{27hf}
f(R,\theta)=\frac{1}{16R^{3/2}\sin\frac{\theta}{2}}~, }
which is harmonic in $\mathbb{R}^4$ except for $\theta=0$. The NS-NS two-form and dilaton are given by,
\bea
{B}_2 &=&-\frac{R^{1/2}}{4}\cos^3\frac{\theta}{2} \sin\chi d\chi\wedge d\xi \nn \\
H_3   &=& -\frac{\cos^3\frac{\theta}{2}}{8 R^{1/2}}     \d R\wedge \sin\chi d\chi\wedge d\xi + \frac{3}{8}R^{1/2} \cos^2\frac{\theta}{2}\sin\frac{\theta}{2} \d\theta\wedge \sin\chi d\chi\wedge d\xi  \\
e^{2{\phi}}&=& 1024\alpha'^3 f~.
\eea
This clearly shows the presence of NS5 branes along the $\mathbb{R}^{1,5}$ directions, located at $\theta=0$ (or alternatively at $u=0$), in accordance with the 
harmonic superposition rule \cite{Youm:1997hw}. However this is not enough to determine the exact position of the branes since they could be anywhere on this half line. Integrating $H_3$ on a spherical shell of radius $R$ gives,
\bea
\int H_3 = \int_{\theta=0}^\pi\int_{\chi=0}^\pi\int_{\xi=0}^{2\pi} dB_2=\pi \sqrt{R} ~.
\eea
The branes are thus smeared along the $\theta=0$ direction, leading to a linear distribution of charge in the transverse space, whose charge density is proportional to $\frac{1}{\sqrt{R}}$ or constant in $\rho$ (recall that at $\theta=0$, $\sqrt{R}=\rho$).

More explicitly the NS5 distribution can be read off of the harmonic function $f$ in \eqref{27hf} as follows.  
First it will be convenient to parameterize the $\mathbb{R}^4$ transverse to the NS5 by introducing the cylindrical coordinates $\vec{r}\in\mathbb{R}^3$, $w:=R\cos\theta$, so that 
$R^2=\vec{r}^{~\!2} + w^2$ and,
\eq{\label{31c}
\d s^2(\mathbb{R}^4) =\d \vec{r}\cdot\d \vec{r} +\d w^2
~.}
The new coordinates $(R,w)$ are related to $(u,\rho)$ by,
\eq{
u^4=\frac12(R-w)~;~~~\alpha^{\prime2}\rho^2=\frac12(R+w)~.
}
Moreover it can easily be verified that the  function $f$  can be represented as an integral over the Green's function for the Laplacian on $\mathbb{R}^4$,
\eq{
f=\frac{\sqrt{2}}{16R\sqrt{R-w}}
=\frac{1}{8\pi}\int_0^\infty \d w' \frac{\sigma(w')}{\vec{r}^{~\!2} +(w-w')^2}
~,}
with linear charge density $\sigma(w)=w^{-\frac12}$ along the half line $w\geq0$. 
This is depicted schematically in diagram (d) of fig.~\ref{fig1}.

An alternative way to  find the charge distribution  is to compute the source for the $H_3$ Bianchi identity,
\begin{equation}
\d H_3= j~.
\end{equation}
However $H_3=\d B_2$ is closed as a form, and we thus need to consider this equation on currents. Indeed  $H_3$ is not defined for $\theta=0$, which is precisely the locus where we expect to find the brane. As a  current, $\d H_3$ acts as a linear form (distribution) on six-forms. Consider a test six-form $\Omega$,
\begin{equation}
\Omega = \omega v_6 ~,
\end{equation}
where $v_6$ is the volume form of $\mathbb{R}^{(1,5)}$.
After integration against $H_3$, the only  components of $\d\Omega$ we need consider are,
\begin{equation}
\d\Omega = \partial_R\omega \d R\wedge v_6 +\partial_\theta\omega \d \theta\wedge v_6+ \cdots~,
\end{equation}
so that,
\begin{equation}
\begin{split}
\d H_3 (\Omega) &= H_3(\d \Omega) \\
              &= \int H_3\wedge\d\Omega \\
              &= -\frac{1}{8}\int \frac{\cos^3\frac{\theta}{2}}{R^{1/2}}\partial_\theta\omega \d R \wedge \sin\chi\d\chi\wedge\d\xi \wedge \d\theta\wedge v_6 \\
              &+\frac{3}{8} \int R^{1/2} \cos^2\frac{\theta}{2}\sin\frac{\theta}{2} \partial_R\omega \d\theta \wedge\sin\chi \d\chi\wedge\d\xi \wedge\d R \wedge v_6~.
\end{split}
\end{equation}
Integrating each term by parts (respectively in $\theta$ and $R$), the derivatives cancel out since $H_3$ is closed as a form. The charge can then be seen in the boundary terms. $R^{1/2}$ vanishes at $R=0$, $\omega$ vanishes at $R\rightarrow \infty$ because it is a test function, and $\cos^3\frac{\theta}{2}$ vanishes at $\theta=\pi$. Note also that at $\theta=0$, $\omega$ cannot depend on $\chi,\xi$. We thus obtain,
\begin{equation}
\begin{split}
\d H_3 (\Omega) &= \frac{1}{8} \int \frac{\omega(\theta=0)}{R^{1/2}} \d R \wedge \sin\chi \d\chi\wedge\d\xi \wedge v_6\\
              &= \frac{\pi}{2} \int \frac{\d R}{R^{1/2}} \wedge \Omega(\theta=0)~.
\end{split}
\end{equation}
From this we can read off the current,
\begin{equation}
j = \frac{1}{16} \frac{\delta(\theta)}{R^{1/2}} \d R \wedge\d\theta \wedge \sin\chi\d\chi\wedge\d\xi~,
\end{equation}
which gives the exact distribution of NS5 charge. This distribution is remarkable since it is entirely created by NATD from an empty flat space. It will be characteristic of the behavior of NATD near a fixed point of the $SU(2)$ isometry. For instance we found the same kind of distribution when looking close to the $\alpha=\pi/2$ singularity in the full dual solution (\ref{NS5metric}).

\subsubsection{The near-horizon limit}\label{sec:23}

The NATD of the near horizon solution is \cite{Sfetsos:2010uq},
\bea \label{NHNATD}
{ds}^2&=&ds^2(AdS_5)+L^2(d\alpha^2+\sin^2\alpha d\theta^2)\nn \\&~&+\frac{1}{4}\Big(\frac{\alpha'^2}{\Xi}d\rho^2+\frac{\Xi^2}{64\alpha'\Delta}\rho^2(d\chi^2+\sin^2\chi d\xi^2)\Big)\nn \\
{B}_2&=&-\frac{\rho^3\Xi}{256\Delta}\sin\chi d\chi\wedge d\xi \nn \\
e^{-2\tilde{\Phi}}&=&\Delta, \quad \Delta =  \frac{\Xi}{64\alpha'^3}(\alpha'^2\rho^2+\Xi^2),\quad \Xi=L^2\cos^2\alpha~,
\eea
and the nonzero RR fluxes are given by,
\bea\label{NHNATDRR}
{F}_2&=&\frac{\Xi}{2\alpha'^{3/2}} L^2\cos\alpha\sin\alpha d\alpha\wedge d\theta\nn \\
{F}_4&=&-\frac{\Xi^2}{512\alpha'^{3/2}\Delta}L^2\rho^3\cos\alpha\sin\alpha\sin\chi d\alpha\wedge d\theta\wedge d\chi\wedge d\xi~.
\eea

{\it Field Theory interpretation of near horizon NATD}

In \cite{Lozano:2016kum} a holographic interpretation of the background (\ref{NHNATD})-(\ref{NHNATDRR}) was proposed.  It was pointed out that the background belongs to a class of Gaiotto-Maldacena geometries  \cite{Gaiotto:2009gz} dual to $\mathcal{N}=2$ superconformal linear quivers with gauge groups of increasing rank. Their argument crucially involved constraining the range of the dual coordinate $\rho$ in quantizing the NS5 brane charge.  Let us briefly summarize the main points of the arguments originally presented in \cite{Lozano:2016kum} and extended to further examples in \cite{Lozano:2016wrs,Itsios:2017cew}. Related examples with flavor branes include \cite{Lozano:2012au,Lozano:2013oma} and \cite{Lozano:2014ata,Lozano:2015cra}.

In the NATD a new set of dual coordinates arise, which we have labeled $(\rho,\chi,\xi)$.  The coordinates $(\chi,\xi)$ are naturally interpreted as compact angles on an $S^2$, i.e. $\chi\in [0,\pi],\ \xi\in [0,2\pi]$.  The question remains how to interpret the $\rho$ coordinate, as NATD currently lacks the global information needed to constrain the dual coordinates.  Using insight from string theory the authors of \cite{Lozano:2016kum} were led to  impose the boundedness of the following quantity,
\bea \label{botrick}
b_0=\frac{1}{4\pi^2\alpha'}\oint_{\Sigma_2}  B_2 \in [0,1]~,
\eea
where in the case of \eqref{NHNATD} $b_0$ is  maximal along $\Sigma_2=[\chi,\xi], \alpha=\frac{\pi}{2}$.  This leads to the coordinate $\rho$ varying in $n\pi$ intervals, i.e. $\rho\in [n\pi,(n+1)\pi]$.  To keep the relation \eqref{botrick} satisfied, a large gauge transformation must be performed on $B_2$ at each $n\pi$ interval, i.e. 
\bea 
B_2\to B_2-n\pi\alpha'\sin\chi d\chi\wedge d\xi~.
\eea
As reviewed in section \ref{sec:235}, 
this has the effect of changing the Page charges:  quantizing $Q_{D6}$ and $Q_{D4}$ by integrating the RR fluxes in \eqref{NHNATDRR} above leads to $Q_{D6}=N_{D6}$ and $Q_{D4}=0$. However under a large gauge transformation of $B_2$, we find $\Delta Q_{D6}=0$ and $\Delta Q_{D4}=nN_{D6}$,  where $Q_{NS5}=N_{NS5}=n$.  Putting all this together suggests  that there are parallel NS5 branes, each located at a $\pi$ interval in $\rho$.  Between each $\pi$ interval $n$ horizontal D4 branes are suspended between them.  That is, as we move towards larger $\rho$, an increasing number of D4 branes appear.  In the field theory interpretation this corresponds to an infinite linear quiver with increasing gauge group ranks.
Interestingly, the field theory analysis of \cite{Lozano:2016kum} suggested that there should be a cutoff to the $\rho$ coordinate in order to terminate the quiver with a flavor brane.
 The intuitive way to see this is to start with parallel NS5 branes and a D6 flavor brane on one of the ends of the array.  When one moves this flavor brane across the NS branes, D4 branes are created across the NS branes via the Hanany-Witten effect \cite{Hanany:1996ie}.  This completion of the quiver corresponds to giving $\rho$ a finite range and it was shown that this is necessary to make sense of the dual field theory as a 4d CFT.\footnote{It was suggested in \cite{Lozano:2016kum} and further considered in \cite{Itsios:2017nou} that the dual field theory could actually be higher dimensional through deconstruction.}

Thus the ``stringy'' picture is consistent with the spatial infinity limit of section \ref{sec:24}, provided we
replace the supergravity approximation of a continuous linear distribution of NS5 branes along a half line, by a grid of localized 
NS5's so that there is one unit of NS5 charge per $\rho\in [n\pi,(n+1)\pi]$ interval.

\section{M2 branes}\label{sec:0}

The M2 brane solutions of eleven-dimensional supergravity can be reduced in various ways in order to obtain 
ten-dimensional IIA D2 brane solutions.  Let us  start from the M2-brane solution in flat space,  
\eq{\spl{\label{1}
ds^2&=H^{-2/3}ds^2(\mathbb{R}^{1,2})+H^{1/3}(\d r^2+r^2\d\Omega_7^2)\\
G&=-\d H^{-1} \wedge \mathrm{vol}_3\\
H&=1+\frac{\hat{Q}}{r^6}
~,}}
where $\hat{Q}$ is a constant related to the number of parallel M2-branes, 
$\d\Omega_7^2$ is the metric of the round seven-sphere, and vol$_3$ is the volume element of $\mathbb{R}^{1,2}$.
We will adopt the parameterization of the metric on $S^7$ given by,
\bea
\d\Omega_7^2&=&\frac{1}{4}\Big( d\mu^2 + \frac{1}{4}\big(\sin^2\mu \omega_i^2 +\lambda^2 (\nu_i+\cos\mu \omega_i)^2\big)\Big), \nonumber \\
\nu_i&=& \sigma_i +\Sigma_i, \qquad \omega_i=\sigma_i -\Sigma_i~,
\eea
where $\mu\in[0,\pi]$, $\sigma_i$  are the left-invariant $SU(2)$ Maurer Cartan one-forms given in \eqref{si}, 
while the $\Sigma_i$ have exactly the same form but with coordinates ($\theta_2,\phi_2,\psi_2$).  We will only treat the round $S^7$ case, i.e. $\lambda=1$.
 In the near-horizon limit, we have $H= \frac{\hat{Q}}{r^6}$ and  the space becomes AdS$_4\times S^7$.

This solution preserves 16 real supercharges (enhanced to 32 in the near-horizon limit), {\it i.e.}~$\mathcal{N}=4$ in four dimensions. 
In (\ref{1}) we have written the flat metric on the space $\mathbb{R}^{8}$ transverse to the M2 as an eight-dimensional cone over the seven-sphere. 
We may replace the base of the cone by any Sasaki-Einstein seven-manifold\footnote{The metric of the Sasaki-Einstein 
manifold must be normalized so that the cone over it is Ricci-flat.}, and still obtain a solution  of eleven-dimensional supergravity. The amount 
of preserved supersymmetry depends on the number of Killing spinors of the Sasaki-Einstein.

Replacing the round sphere metric $\d\Omega_7^2$ by  the $Y^{p,q}(B_4)$ metric of \cite{Gauntlett:2004hh,Martelli:2004wu}, reduces supersymmetry to $\mathcal{N}=1$ in four dimensions, 
enhanced to $\mathcal{N}=2$ in the near-horizon limit.  
After a change of coordinates to bring us to the conventions 
of \cite{PandoZayas:2017ier}, the metric reads,
\eq{\label{se7}
\d s^2(Y_7)= \frac14 \d s^2(M_6)
+w(\theta)\left[
\d\alpha+f(\theta)(\d\psi+A)
\right]^2
~,
}
for some functions $w$, $f$  of $\theta$ that will be specified below, where $\d s^2(M_6)$ is the metric of the $S^2(B_4)$ bundle,
\begin{equation}\label{2}
\d s^2(M_6) =  \d s^2(B_4)+\frac{1}{(1+\cos^2\theta)^2}\d\theta^2+\sin^2\theta(\d\psi+{A})^2~,
\end{equation}
with $\theta\in\left[0,\pi\right]$, $\psi\in\left[0,\pi\right]$ the coordinates of the $S^2$ fiber; the connection $A$ is a one-form on the base $B_4$ obeying,
\eq{\label{ccon}
\d{} A = J
~,
}
with $J$ the K\"{a}hler form of $B_4$.  
Later we will consider the special case $B_4=\cp2$ for concreteness and in order to perform an SU(2) NATD.

The corresponding eleven-dimensional solution reads,
\eq{\spl{\label{3}
ds^2&=H^{-2/3}ds^2(\mathbb{R}^{1,2})+H^{1/3}\left(\d r^2+\frac14 r^2\d s^2(M_6)\right)+r^2 H^{1/3}w(\theta)
(\d\alpha+A')^2\\
G&=-\d H^{-1} \wedge \mathrm{vol}_3\\
H&=1+\frac{\hat{Q}}{r^6}
~,}}
where we have set $A':=f(\theta)(\d\psi+A)$.

\subsection{Brane configuration and charges}

We expect the M2 branes to lie along the $\mathbb{R}^{1,2}$ directions. The transverse space would then be $\mathbb{R}^8$ or the cone over $Y^{p,q}$ depending on the choice of 7-dimensionnal space. In both cases we find:
\begin{equation}
\star G= -6\hat{Q} v_7
\end{equation}
However since the 7-dimensional cycle collapses in the transverse space when $r=0$, $\star G$ is not closed and:
\begin{equation}
\d\star G = -6\hat{Q} \delta(r) \d r\wedge v_7
\end{equation}
We can also compute the M2 brane charge, which is defined by:
\bea
Q_{M2}=\frac{1}{2\kappa^2_{11}T_{M2}}\int \star ~\!G=N_{M2}~,
\eea
with the M2 brane tension given by $T_{M2}=\frac{2\pi}{(2\pi l_p)^3}$ and $2\kappa^2_{11}=(2\pi)^8l_p^9$,  
where the Planck length is given by  $l_p=g_s^{1/3}\sqrt{\alpha'}$. 
For example, in the $Y^{p,q}$ case, we compute:
\bea
Q_{M2}&=&-\frac{1}{2\kappa_{11}^2T_{M2}}\frac{27\hat{Q}}{256}\int_0^{2\pi}d\alpha \int_0^{\pi}d\theta \frac{\sin\theta}{a(\theta)^{3/2}} \int_{S^3}d\Omega_3\int_0^{\pi}d\psi\int_0^{\frac{\pi}{2}}d\mu\sin^3\mu,\nn \\ 
&=&-\frac{27\hat{Q}}{4096\pi^2l_p^6}=N_{M2}~.
\eea
This relates the constant in the harmonic function to the number of M2 branes,
\bea
\hat{Q}=\frac{4096}{27}\pi^2l_p^6N_{M2}~.
\eea
We will now proceed to track the M2, first through dimensional reduction, then through NATD.

\section{Supersymmetric D2 from reduction on $Y^{p,q}$}\label{sec:5}

Here and in the following section we will  need to make the choice  ${B}_4=\mathbb{CP}^2$ so that a non-abelian SU(2) isometry is manifest in the metric acting on  
the $\sigma_i$, 
\bea
\label{CP2}
ds^2(\mathbb{CP}^2)=3 \big[d\mu^2+\frac{1}{4}\sin^2\mu(\sigma_1^2+\sigma_2^2+\cos^2\mu \sigma_3^2)\big]~,
\eea
where $\mu\in[0,\frac{\pi}{2}]$, and the $\sigma_i$ are given in \eqref{si}.

Reducing the M2 brane solution \eqref{3} to IIA on the circle parameterized by $\alpha$ preserves supersymmetry, as  will be explicitly verified in section \ref{sec:susydw}. Let us set,
\eq{\spl{ \label{4}
e^{-2\phi/3}\d s^2_{\mathrm{A}}&=  H^{-2/3}ds^2(\mathbb{R}^{1,2})+H^{1/3}\left(\d r^2+\frac14 r^2\d s^2(M_6)\right) \\
e^{4\phi/3}&= \frac{r^2}{l_p^2} H^{1/3}w(\theta)~,
}}
so that upon reduction to ten dimensions $\d s^2_{\mathrm{A}}$ and the function $\phi(r,\theta)$ are identified with the IIA string-frame metric and dilaton respectively. 
Moreover the nonvanishing fluxes of the solution are given by,
\eq{\label{5}
F_2=l_p\d A'~;~~~
F_4=-\d H^{-1} \wedge \mathrm{vol}_3~,}
where $A'$ was given below \eqref{3}. 
$F_2$ carries a magnetic charge, but we will not interpret it as coming from a brane. Since this charge is not related to the M2 charge, but rather to the dimensional reduction, we will say that the flux is only geometric, and it will not be of interest here.
The flux $F_4$ on the other hand carries an electric charge and is sourced by a stack of parallel D2 branes filling $\mathbb{R}^{1,2}$ and placed at $r=0$ in the transverse space. Note that the $H$ function is inherited from the M2 solution, so that it does not need to be harmonic in the new transverse space. We also inherit the usual parameters for a brane solution, which allow us to define the near-horizon and spatial infinity limit.

We can then obtain the explicit form of the functions $w(\theta)$, $f(\theta)$ by taking the near-horizon limit ($H=\frac{\hat{Q}}{r^6}$)
of (\ref{4}), (\ref{5}) and comparing with \cite{PandoZayas:2017ier}:
\eq{\spl{\label{6}
\d s^2_{\mathrm{A}}&=
 \frac14 \hat{Q}^{1/2}\sqrt{w(\theta)}\left(  ds^2(\mathrm{AdS}_{4})+\d s^2(M_6)\right)\\
e^{4\phi/3}&=  \hat{Q}^{1/3}w(\theta)~;~~~F_2=l_p\d\left[ f(\theta)(\d\psi+A)\right]
~,}}
where $ds^2(\mathrm{AdS}_{4})$ is the metric of an $\mathrm{AdS}_{4}$ space of unit radius, so that 
its scalar curvature is normalized to $R=-12$. 
Comparing with (2.22), (2.23), (2.24) of \cite{PandoZayas:2017ier} we read off,
\eq{\label{7}
w(\theta)= \frac{g_s^2~\!e^{4A_0}}{8(1+\cos^2\theta)}
~;~~~f(\theta)=\frac{\cos\theta}{2\sqrt{w(\theta)}}~;~~~\hat{Q}=\frac{64}{g_s^2}
~.}
To summarize, the ten-dimensional D2-brane solution is given by \eqref{4}, \eqref{5}, 
%
%
where $\d s^2(M_6)$ is given in (\ref{2}), $H$ is given in (\ref{3}) and $f$, $w$ are given in (\ref{7}). 
In the near-horizon limit the metric becomes 
a warped $\mathrm{AdS}_{4}\times M_6$ product, {cf.}~(\ref{6}).

At spatial infinity ($H=1$) the metric becomes 
a warped product $\mathbb{R}^{1,2}\times C(M_6)$,
\eq{\spl{\label{8}
ds^2_{\mathrm{A}}= \frac{r}{l_p}\sqrt{w(\theta)} \left(\d s^2(\mathbb{R}^{1,2})+ \d r^2+\frac14  r^2\d s^2(M_6)\right)
~,}}
where $C(M_6)$ is the metric cone over $M_6$, while the remaining fields are given by,
\eq{\spl{\label{9}
e^{4\phi/3}&= \frac{r^2}{l_p^2}  w(\theta) \\
F_2&= l_p\d\left[ f(\theta)(\d\psi+A)\right]\\
F_4&=0
~.
}}
It can be verified that this is an exact supergravity solution in its own right. Contrary to the case of the D3 brane, here spacetime is neither flat nor empty at spatial infinity. 

The solution \eqref{4}, \eqref{5} describes D2 branes with worldvolume along the 
$\mathbb{R}^{1,2}$, as inherited from the M2 solution. Looking at $F_4$, we find:
\begin{equation}
\star F_4 = -\frac{3\hat{Q}}{32 l_p}\sqrt{w(\theta)} v_6
\end{equation}
However, contrary to the standard brane configurations (such as the D3 and M2 presented previously), the probe interpretation is not straightforward. In order to understand this configuration we take the transverse space to be the cone over $M_6$. There the 6-cycle collapses at $r=0$, so that:
\begin{equation}
\d\star F_4 = -\frac{3\hat{Q}}{32 l_p}\sqrt{w(\theta)} \delta(r) \d r\wedge v_6
\end{equation}
Here again this equation must be considered on the transverse space. The D2 background is a genuine solution of IIA supergravity, in which the 6-cycle does not collapse anymore. Then $\star F_4$ is closed, as required by the equations of motion, and the brane is not visible. We can however compute the brane charge, which  requires the D-brane tension $T_{Dp}^{-1}=((2\pi)^p\alpha'^{\frac{(p+1)}{2}})$ and $2\kappa_{10}^2=(2\pi)^7\alpha'^4$. 
We obtain,
\bea
Q_{D2}=\frac{27\tilde{Q}}{8192\pi^5l_p\alpha'^{5/2}}\int_0^{\pi}d\psi \int_0^{\pi}d\theta \frac{\sin\theta}{a(\theta)^{3/2}} \int_{S^3}d\Omega_3\int_0^{\frac{\pi}{2}}d\mu\sin^3\mu\cos\mu,\nn 
\eea
The flux quantization condition $Q_{D2}=N_{D2}$ then leads to
\bea\label{QtN0}
\hat{Q}=\frac{4096}{27}\pi^2l_p\alpha'^{5/2}N_{D2}~.
\eea
Note  that there are no D6 branes associated with the $F_2$ flux. Indeed in the present case  spacetime is smooth\footnote{The geometry and topology of the M-theory reduction along the $\alpha$-cycle is discussed in detail in \cite{Martelli:2008rt}.} and the metric singularity expected in the vicinity of a D6 is absent. As we will see  in section \ref{sec:3}, this is in contrast to the case of the D2 brane coming from the reduction of M-theory on $S^7$. Similarly one sees that there are no $D4$ branes sourced by the $F_4$ flux.

\subsection{The NATD}\label{sec:51}


The NATD of the supersymmetric D2 brane is obtained by an SU(2) action on the $\sigma_i$, cf.~\eqref{CP2}. 
The NS-NS sector reads,
\bea \label{MasslessNS}
\hat{ds}^2&=& \frac{r}{l_p}\sqrt{w(\theta)}H^{-1/2}ds^2(\mathbb{R}^{1,2})+\Lambda^2\big(\frac{4}{r^2}dr^2+ 3 d\mu^2+\frac{1}{(1+\cos^2\theta)^2}d\theta^2 \nn \\&~&+\frac{4}{Q}\sin^2\theta\cos^2\mu d\psi^2\big)+\frac{3\alpha'^2\Xi}{4M} [d(\rho\sin\chi)]^2\nn \\&~&+\frac{81}{4096\alpha'\Delta}\bigg[ \frac{ \Xi^2\rho^2\sin^2\chi}{Q}(d\xi\psi)^2+\frac{1}{M}\bigg(\alpha'^2\rho^2\cos\chi d\rho+4\Xi^2 d(\rho\cos\chi)\bigg)^2\bigg] \nn \\
\hat{B}_2&=&\frac{81\rho^2 \Xi \sin\chi}{8192 Q \Delta}d\xi\psi\wedge d\rho\chi +\frac{3\alpha'\sin^2\theta}{2Q}d(\rho\cos\chi)\wedge d\psi \nn \\
e^{-2\hat{\phi}}&=& e^{-2\phi}\Delta,\quad \Delta=\frac{27\Xi}{1024\alpha'^3}\big(4\Xi^2 Q + \alpha'^2\rho^2  K\big), \quad \Xi=\sin^2\mu\Lambda^2, \quad \Lambda=\frac{1}{2}e^{\phi/3}rH^{1/6}~,
\eea
where we have defined the following one-forms,
\bea\label{4.2}
d\xi\psi&=& \big(Q d\xi-4\sin^2\theta d\psi\big)\nn \\
d\rho\chi&=&\Big(\rho  K d\chi +\cos\chi \sin\chi(Q-4) d\rho\Big)\nn \\
d\theta\mu&=&\big(f'(\theta)\sin\mu d\theta+2\cos\mu f(\theta) d\mu\big)~,
\eea
and included the following definitions,
\bea \label{QKMa}
Q&=&4\cos^2\mu+3\sin^2\mu\sin^2\theta\nn \\
K&=&Q \cos^2\chi+4\sin^2\chi\nn \\
M&=&\alpha'^2\rho^2\cos^2\chi+4\Xi^2~.
\eea
The RR sector is given by
\bea \label{MasslessRR}
\hat{F}_1&=&\frac{9l_p}{32\sqrt{\alpha'}}\sin\mu\bigg[ f(\theta)\sin\mu d(\rho\cos\chi)  -\rho\cos\chi d\theta\mu\bigg]\nn \\
\hat{F}_3&=&-\bigg(\frac{9l_p\sqrt{\alpha'}\rho \cos^2\mu f'(\theta) }{16Q}d\rho+\frac{9l_p^2\Lambda^6H'\cos\mu\sin^3\mu\sin\theta}{4r^2\alpha'^{3/2}H^{3/2}w(\theta)a(\theta)} d\mu\bigg)\wedge d\theta\wedge d\psi \nn \\&~& 
+\frac{729l_p\rho^3\sin^3\mu\Lambda^2}{262144\sqrt{\alpha'} Q\Delta} \bigg[-\cos\chi\sin\chi\sin\mu Q d\rho\wedge d\chi\wedge d\xi\psi \nn \\&~& +2\cos\mu f(\theta)Q (\sin^2\chi d\xi-\cos^2\chi\sin^2\mu d\psi)\wedge d\mu \nn \\&~& -\sin\mu\sin^2\chi f'(\theta) d\theta \wedge d\xi\psi\bigg]\wedge d\rho \nn \nn \\&~& +\frac{729l_p\rho\sin^7\mu\Lambda^6}{65536\alpha'^{5/2} \Delta} \big[\sin\chi d\theta\mu\wedge d(\rho\sin\chi)\wedge d\xi\psi\nn \\&~& -8\cos\mu\sin^2\theta f(\theta) d\mu\wedge d\rho\wedge d\psi\big] 
\nn \\
\hat{F}_5&=&\frac{9l_p^2\sqrt{\alpha'}H'\rho}{64r^2H^{3/2}w(\theta)} v_4\wedge d\rho \nn \\&~&+\frac{9l_p\Lambda^2\sin^3\mu}{16\alpha'^{3/2} \sin\theta a(\theta)}v_4\wedge\Big(2f(\theta)\sin^2\theta\sin\mu d\theta+a(\theta)^2\cos\mu f'(\theta) d\mu  \Big) 
\nn \\&~&+\frac{729l_p \Lambda^6\rho^2\cos\mu\sin^5\mu\sin\chi}{65536\alpha'^{3/2} r^2H^{3/2}w(\theta)a(\theta)\Delta}\bigg[2l_p\Lambda^2 H'\sin\theta d\theta\wedge d\mu\wedge d\rho\chi \nn \\&~&-r^2H^{3/2}w(\theta)a(\theta)\sin\mu\big(3f(\theta)\sin^2\theta \sin\mu d\mu\nn \\&~& -2\cos\mu f'(\theta) d\theta\big)\wedge d\rho\wedge d\chi \bigg]\wedge d\xi\wedge d\psi~,
\eea
where $a(\theta)=2(1+\cos^2\theta)$ and
$v_4=-\frac{r^2 w(\theta)}{l_p^2\sqrt{H}}dr\wedge dx_0\wedge dx_1\wedge dx_2$.

\subsubsection{Brane configuration and charges}\label{sec:52}

The D-brane background before the NATD was a D2 brane solution, therefore we expect to see the presence of D3, D5, and NS5 branes from the general lore Dp$\rightarrow$D(p+1)-D(p+3)-NS5. We will follow the same strategy as in section \ref{sec:235} to better understand the brane configuration.

We first compute the NS5 charge. In the same spirit as in the D3 brane example, cf.~\eqref{B2H3D3}, we integrate $H_3$ along the cycle $(\Sigma_3[\rho,\chi,\xi],\mu=0)$, on which $H_3$ simplifies as,
\bea
H_3&=&\frac{3}{8}\alpha' \sin\chi d\xi\wedge d\chi\wedge d\rho~.
\eea
We get,
\begin{equation}
Q_{NS5} = \frac{1}{2\kappa_{10}^2 T_{NS5}} \int H_3 = \frac{3}{8\pi}\rho_0~,
\end{equation}
where we cut off the integration at $\rho=\rho_0$. For the charge to be correctly quantized, we need $\rho_0=\frac{8n\pi}{3}$. This is compatible with the condition of boundedness of $b_0$ given in (\ref{botrick}). Modulo a large gauge transformation on $B_2$, this condition is satisfied if the range of $\rho$ is taken to be $[\frac{8(n-1)\pi}{3},\frac{8n\pi}{3}]$. Once again we can see that there is, at least from the supergravity point of view, a continuous distribution of charge at the singularity created by NATD (here $\mu=0$). This distribution is smeared along the $\rho$ direction and is constant in $\rho$. As was the case in section \ref{sec:235}, this can be seen directly in the metric by zooming in at the singularity. Close to $\mu=0$ and after making the substitution $\nu=\mu^2$, the metric becomes,
\bea
ds^2_{\mu\to 0}&=&\frac{r}{2l_p\sqrt{H(r)}\sqrt{a(\theta)}}\Big[ds^2(\mathbb{R}^{1,2})+H(r) \big(dr^2+ r^2\big(\frac{1}{a(\theta)^2}d\theta^2 + \frac{1}{4}\sin^2\theta d\psi^2 \big)\big)\Big]\nn\\
&+&\frac{1}{\nu}\Big[\frac{3}{32l_pr^3\sqrt{H(r)}\sqrt{a(\theta)}}\bigg(16l_p^2\alpha'^2a(\theta)^2 d\rho^2+r^6H(r)\Big[d\nu^2+\nu^2\big(d\chi^2\nn \\&+&\sin^2\chi d\xi(d\xi-2\sin^2\theta d\psi)\big)\Big]\bigg)\Big]
\label{72ac}
~,\eea
where $\nu^{-1}$ is the harmonic function in the transverse space for NS5 branes along the $(\mathbb{R}^{1,2},r,\theta,\psi)$ directions.

For the D-branes we need to consider the Page forms, given by,
\bea
\tilde{F}_3&=&\frac{9l_p\sqrt{\alpha'}\rho}{256}\big[4f'(\theta)d\theta\wedge d\rho\wedge d\psi-3\sin\chi \sin\mu d\theta\mu\wedge d(\rho\sin\chi)\wedge d\xi\big]\nn \\&~&
-\frac{27\hat{Q}\sqrt{w(\theta)}}{128 l_p\alpha'^{3/2}a(\theta)}\cos\mu\sin^3\mu\sin\theta  d\theta\wedge d\mu\wedge d\psi \nn \\
\tilde{F}_5&=&-\frac{27 \hat{Q} l_p^2\sqrt{\alpha'}}{32r^9 H^{3/2}w(\theta)}\rho d\rho\wedge v_4+\frac{9r^3\sqrt{H}\sqrt{w(\theta)}}{64\alpha'^{3/2}a(\theta)\sin\theta}\sin^3\mu (2f(\theta)\sin^2\theta \sin\mu d\theta \nn \\&~&+a(\theta)^2\cos\mu f'(\theta) d\mu)\wedge v_4\nn \\&~& +\frac{27l_p\alpha'^{3/2}}{512}\rho^2\sin\chi f'(\theta) d\theta \wedge d\rho \wedge d\chi\wedge d\xi\wedge d\psi~,
\eea
where $\tilde{F}_1=\hat{F}_1$ given in \eqref{MasslessRR} is unchanged.

We can readily see that the Page forms have two contributions: one coming from the geometric flux $F_2$ and the other one from $F_4$.\footnote{Recall that NATD acts linearly on the RR fluxes so we can isolate each contribution.} Since we want to trace the fate of the M2 branes we will consider the first part as geometric fluxes and focus on the second. The relevant components are thus those proportional to $\hat{Q}$ (i.e. those that vanish when there is no M2).

Ignoring the geometric fluxes, the only non-vanishing Page charge is $Q_{D5}$, which can be found by integrating $\tilde{F}_3$. Namely, we keep only the $(\theta,\mu,\psi)$ term,
\bea
Q_{D5}=\frac{1}{2\kappa_{10}^2T_{D5}}\frac{27\hat{Q}}{128 l_p\alpha'^{3/2}}\int_0^{4\pi}d\psi\int_0^{\pi}\frac{\sin\theta\sqrt{w(\theta)}}{a(\theta)}d\theta\int_0^{\frac{\pi}{2}}\cos\mu\sin^3\mu d\mu=N_{D5}~.
\eea
The quantization condition of $Q_{D5}$ then leads to a relation between the constant in the harmonic function and the number of D5 branes,
\bea \label{deltaQD5}
\hat{Q}=\frac{2048}{27}\pi l_p\alpha'^{5/2}N_{D5}~.
\eea
However as it was pointed out in the D3 example, the page charges depend on the choice of $B_2$, and may change under a large gauge transformation. Here under a large gauge transformation given by,
\begin{equation}\label{deltaB2}
\Delta B_2 = - n \pi \alpha' \sin\chi\d\chi\wedge\d\xi~,
\end{equation}
$Q_{D3}$ receives a new contribution,
\bea
\Delta Q_{D3}&=&\int -\Delta B_2 \wedge F_3    \\
             &=&\frac{1}{2T_{D3}\kappa^2}\frac{27n\pi\hat{Q}}{128 l_p\sqrt{\alpha'}}\int_0^{\frac{\pi}{2}}\cos\mu\sin^3\mu d\mu\int_{0}^{\pi}\frac{\sin\theta\sqrt{w(\theta)}}{a(\theta)}d\theta\int_0^{4\pi}d\psi  \int_0^{\pi}\sin\chi d\chi \int_0^{2\pi}d\xi
             ~.\nn
\eea
Evaluating this and comparing to  (\ref{deltaQD5}), we then find, 
\bea
\Delta Q_{D3}=n N_{D5}.
\eea
This is analogous to the relation found in  (\ref{deltaQD4}) above.

Assuming that the $r$ coordinate still describes the radius of the cycles wrapped by the RR-fluxes in the transverse space of the D-branes, the brane configuration is given by the table below.
\begin{center}
	\begin{tabular}{ |c|ccc|ccccccc|} 
		\hline
		& 0&1&2&$ r$&$\mu$&$\theta$&$\psi$ &$\rho$&$\chi$&$\xi$ \\
		\hline
		$NS5$ & $\times$ &$\times$ &$\times$ &$\times$ & & $\times$ &$\times$ & & & \\ 
		$D5$  & $\times$&$\times$&$\times$& & & & &$\times$ &$\times$ & $\times$\\ 
		$D3$ & $\times$&$\times$& $\times$ & &  & & &$\times$ & & \\ 
		\hline
	\end{tabular}
\end{center}

As we will find in additional examples throughout this paper, this relationship seems to be universal for D-brane backgrounds generated by SU(2) non-Abelian T-duality.  When a Dp-brane background is transformed, the D(p+3) brane charges are easily found from integrating the appropriate term in the Page form.  The D(p+1) brane charges then are found from restricting $B_2$ to a cycle containing ($\chi,\xi$) cycle, performing a large gauge transformation, computing the change in the Page form under this transformation, and finally integrating to obtain $\Delta Q_{D(p+1)}$.  This results in the general relation $\Delta Q_{D(p+1)}=nQ_{D(p+3)}$. The new D2 brane examples presented in this paper are not only distinct from the original D3 brane example where this relation was proposed, but also highly nontrivial.


\subsubsection{The spatial infinity limit}\label{sec:54}


The spatial infinity limit of the supersymmetric D2 brane NATD solution  \eqref{MasslessNS}-\eqref{MasslessRR} is given by,
\bea
\tilde{ds}^2&=& \frac{r}{l_p}\sqrt{w(\theta)}\bigg[(\mathbb{R}^{1,2})+dr^2+\frac{r^2}{4}\big(3 d\mu^2+\frac{1}{(1+\cos^2\theta)^2}d\theta^2 +\frac{4}{Q}\sin^2\theta\cos^2\mu d\psi^2\big)\nn \\&~&+\frac{3l_p\alpha'^2\sqrt{w(\theta)}\Xi}{4M}[d(\rho\sin\chi)]^2\bigg]+\frac{81}{16384l_p^2\alpha'\Delta}\bigg[\frac{w(\theta)\Xi^2\rho^2\sin^2\chi}{4Q}(d\xi\psi)^2\nn \\&~&+\frac{1}{M}\bigg(4l_p^2\alpha'^2\rho^2\cos\chi d\rho+w(\theta)\Xi^2 d(\rho\cos\chi)\bigg)^2\bigg], \nn \\
\hat{B}_2&=&\frac{81\sqrt{w(\theta)}\Xi  \rho^2 \sin\chi}{32768l_p Q \Delta}d\xi\psi\wedge d\rho\chi +\frac{3\alpha'\sin^2\theta}{2Q}d(\rho\cos\chi)\wedge d\psi,\nn \\
e^{-2\hat{\phi}}&=& e^{-2\phi}\Delta,\quad \Delta=\frac{27\sqrt{w(\theta)}\Xi}{16384l_p^3\alpha'^3}\big(4l_p^2\alpha'^2 \rho^2  K+w(\theta) Q\Xi^2 \big), \quad \Xi=r^3\sin^2\mu,
 \label{80ac}
\eea
where we have defined the following one-forms,
\bea
d\xi\psi&=& \big(Q d\xi-4\sin^2\theta d\psi\big),\nn \\
d\rho\chi&=&\Big(\rho  K d\chi +\cos\chi \sin\chi(Q-4) d\rho\Big),\nn \\
d\theta\mu&=&\big(f'(\theta)\sin\mu d\theta-2\cos\mu f(\theta) d\mu\big),
\eea
with $Q,K$ defined in \eqref{QKMa} and 
$M=4l_p^2\alpha'^2\rho^2\cos^2\chi+w(\theta)\Xi^2$.
The RR sector is given by
\bea \label{MasslessRRasym}
\hat{F}_1&=&\frac{9l_p}{32r^3\sqrt{\alpha'}}\bigg[ f(\theta)\Xi d(\rho\cos\chi)  -\rho\cos\chi d\theta\mu\bigg],\nn \\
\hat{F}_3&=&\frac{9l_p\sqrt{\alpha'}}{16Q}\rho\cos^2\mu f'(\theta) d\theta\wedge d\rho\wedge d\psi\nn \\&~&
+\frac{729r^3\sqrt{w(\theta)}\rho^3\sin^3\mu }{1048576\sqrt{\alpha'} Q\Delta} \bigg[-\cos\chi\sin\chi\sin\mu Q d\rho\wedge d\chi\wedge d\xi\psi \nn \\&~& +2\cos\mu f(\theta)Q (\sin^2\chi d\xi-\cos^2\chi\sin^2\mu d\psi)\wedge d\mu \nn \\&~& -\sin\mu\sin^2\chi f'(\theta) d\theta \wedge d\xi\psi\bigg]\wedge d\rho \nn \\&~&+\frac{729r^9w(\theta)^{3/2}\rho\sin^7\mu}{4194304l_p^2\alpha'^{5/2} \Delta} \big[\sin\chi d\theta\mu\wedge d(\rho\sin\chi)\wedge d\xi\psi\nn \\&~& -8\cos\mu\sin^2\theta f(\theta) d\mu\wedge d\rho\wedge d\psi\big] 
\nn \\
\hat{F}_5&=&\frac{9r^3\sqrt{w(\theta)}\sin^3\mu}{64\alpha'^{3/2} \sin\theta a(\theta)}v_4\wedge\Big(2f(\theta)\sin^2\theta\sin\mu d\theta+a(\theta)^2\cos\mu f'(\theta) d\mu  \Big) 
\nn \\&~&\frac{729 r^9w(\theta)^{3/2}\rho^2\cos\mu\sin^6\mu\sin\chi}{4194304l_p^2\alpha'^{3/2} \Delta}\big(3f(\theta)\sin^2\theta \sin\mu d\mu\nn \\&~& +2\cos\mu f'(\theta) d\theta\big)\wedge d\rho\wedge d\chi \wedge d\xi\wedge d\psi~.
\eea
As was already the case for the D2 solution, the spatial infinity is neither flat nor empty. In a probe interpretation, this configuration could be interpreted as the space in which the branes are inserted. As can be seen from \eqref{80ac}, it is a foliation over the ``radial'' coordinate $r$ with  leaves of the form of a warped product $\mathbb{R}^{1,2}\times\tilde{M}_6$. 
At fixed $r$, the space $\tilde{M}_6$ can be thought of as a fibration of the space $\tilde{N}_3$ parameterized by $(\rho,\chi,\xi)$ fibered over the base $\tilde{M}_3$ parameterized 
by $(\mu,\theta,\psi)$. The topology of  $\tilde{M}_3$  can be deduced from the line element,
\eq{\label{82bd}
\d s^2(\tilde{M}_3):=3 \d\mu^2+\frac{1}{(1+\cos^2\theta)^2}\d\theta^2 +\frac{4\cos^2\mu}{Q}\sin^2\theta \d\psi^2
~,}
and it  is that of an $S^2$ parameterized by $(\theta, \psi)$, fibered over the interval parameterized by $\mu$. 
Indeed at fixed $\mu$, $\d s^2(\tilde{M}_3)$ is of the form $g(\theta)\d\theta^2+h(\theta)\d\psi^2$, for some positive functions $g$, $h$ of $\theta$. 
This has the topology of a  circle parameterized by $\psi$, fibered over the interval parameterized by $\theta$. Moreover at the endpoints of the 
$\theta$-interval,  $Q$ is equal to $4\cos^2\mu$ and the metric becomes $\frac14 \d\theta^2+\sin^2\theta\d\psi^2$ in the vicinity of  $\theta=0,\pi$. This is smooth given that the period of $\psi$ is equal to $\pi$. 
In other words  the $\psi$-circle degenerates 
to a point at the endpoints of the $\theta$-interval so that the total space remains smooth. 
We thus obtain the topology  of an $S^2$, as advertised.

The range of the coordinate $\rho$ was constrained by flux quantization to be the interval specified in section \ref{sec:52}. Moreover over a fixed base point 
$(\mu,\theta,\psi)\in\tilde{M}_3$, the coordinates   $(\chi,\xi)$ parameterize a smooth $S^2$ 
provided we take $\xi\in[0,2\pi]$, $\chi\in[0,\pi]$. This can already be seen from the geometry near the location of the NS5 branes, cf.~\eqref{72ac}. More generally 
the geometry of the  $\tilde{N}_3$ fiber 
over a fixed point in $\tilde{M}_3$ is rather complicated, as can be seen from \eqref{80ac}. Topologically it is an $S^2$   parameterized by $(\chi,\xi)$ fibered over 
the interval parameterized by $\rho$. Indeed at constant $\rho$ the line element of $\tilde{N}_3$ is proportional to,
\eq{\label{83bd}
\frac{3l_p\alpha'^2\sqrt{w(\theta)}\Xi}{4M}\big(\cos^2\chi+
\frac{27\sqrt{w(\theta)}\Xi}{2^{12}l_p^3\alpha^{\prime3}\Delta}\sin^2\chi
\big)\d\chi^2
+\frac{81w(\theta)\Xi^2\sin^2\chi}{2^{16}l_p^2\alpha'\Delta Q}  d\xi^2
~,
}
which is a circle parameterized by $\xi$ fibered over the interval parameterized by $\chi$. Moreover it can be seen that 
near the endpoints of the interval $\chi=0,\pi$ the line element above reduces to,
\eq{\label{84bd}
\frac{3l_p\alpha'^2\sqrt{w(\theta)}\Xi}{4(\alpha'^2\rho^2\cos^2\chi+4\Xi^2)}\big(\d\chi^2+
\sin^2\chi\d\xi^2
\big)
~,
}
so  that the $S^2$ parameterized by $(\xi,\chi)$  is smooth for the ranges given above.

We have thus been able to specify the ranges of all coordinates parameterizing the NATD space.
Once this result has been established for the leaf of the $r$-foliation at spatial infinity, it remains 
valid for finite $r$ and applies also to the full interpolating solution \eqref{MasslessNS}. 
In particular the smoothness of the $S^2$ parameterized by $(\theta,\psi)$ is shown by the same argument following \eqref{82bd}. 
The  smoothness of the $S^2$ parameterized by $(\xi,\chi)$ also follows as above, upon modifying \eqref{83bd}, \eqref{84bd} 
to account for the interpolating metric \eqref{MasslessNS}. 

The near-horizon limit is obtained by substituting $H\rightarrow\tfrac{\hat{Q}}{r^6}$ in \eqref{MasslessNS}. As is clear from the previous 
analysis, the general structure of the leaves of the $r$-foliation described above remains unchanged. Moreover the $\mathbb{R}^{1,2}$ space combines 
with the radial coordinate to form an AdS$_4$ factor exactly as before the NATD.

\section{Non-supersymmetric D2 from reduction on $Y^{p,q}$}\label{sec:4}

We will now  reduce along the ``obvious'' Sasaki-Einstein $S^1$ cycle, thereby completely breaking supersymmetry. 
Let us rewrite   the $Y^{p,q}(B_4)$ metric (\ref{se7}) as follows,
\eq{\label{se7r}
\d s^2(Y_7)=  \frac14\left( \d s^2(\tilde{M}_6)
+(\d\psi+\tilde{A})^2\right)
~,
}
where the base $\tilde{M}_6$ is topologically an $\mathbb{CP}^2\times S^2$ with metric  given by,
\begin{equation}\label{2t}
\d s^2(\tilde{M}_6) =  \d s^2(B_4)+\frac{1}{(1+\cos^2\theta)^2}\d\theta^2 +4w(\theta)\sin^2\!\theta~\!\d\alpha^2~,
\end{equation}
and we have defined,
\eq{\label{bdef}
\tilde{A}:=A+2\sqrt{w(\theta)}\cos\theta\d\alpha
~.}
Note that, as follows from (\ref{7}), for the $S^2$ parameterized by $(\theta,\alpha)$ to be smooth $\alpha$ must have period $2\pi/(g_s~\!e^{2A_0})$. 
Alternatively we may redefine $\alpha\rightarrow g_s~\!e^{2A_0}\alpha$, so that $\alpha\in[0,2\pi]$. In terms of the redefined coordinate,
\begin{equation}\label{2tn}
\d s^2(\tilde{M}_6) =  \d s^2(B_4)+\frac{1}{(1+\cos^2\theta)^2}\d\theta^2 +\frac{\sin^2\theta}{2(1+\cos^2\theta)}\d\alpha^2~.
\end{equation}
The corresponding eleven-dimensional solution reads,
\eq{\spl{\label{3nons}
ds^2&=H^{-2/3}ds^2(\mathbb{R}^{1,2})+H^{1/3}\left(\d r^2+\frac14 r^2\d s^2(\tilde{M}_6)\right)+\frac14 r^2 H^{1/3} 
(\d\psi+\tilde{A})^2\\
G&=-\d H^{-1} \wedge \mathrm{vol}_3\\
H&=1+\frac{\hat{Q}}{r^6}
~.}}
Reducing along the $S^1$ cycle parameterized by $\psi$ results in a non-supersymmetric ten-dimensional D2-brane solution  given by,
\eq{\spl{\label{nons}
ds^2_{\mathrm{A}}&=e^{2\phi/3}\left(H^{-2/3}\d s^2(\mathbb{R}^{1,2})+H^{1/3}\d r^2+\frac14 H^{1/3}r^2\d s^2(\tilde{M}_6)\right)\\
e^{4\phi/3}&=  \frac{r^2}{4l_p^2}   H^{1/3}\\
F_2&=l_p\d \tilde{A}\\
F_4&=- \d H^{-1} \wedge \mathrm{vol}_3
~.
}}
In the spatial infinity limit, $H=1$,
the metric reduces to,
\eq{\spl{\label{asymN0}
ds^2_{\mathrm{A}}= \frac{r}{2l_p} \left(\d s^2(\mathbb{R}^{1,2})+ \d r^2+\frac14  r^2\d s^2(\tilde{M}_6)\right)
~,}}
while the remaining fields reduce to,
\eq{\spl{\label{asymRRNO}
e^{4\phi/3}&=  \frac{r^2}{4l_p^2} \\
F_2&= l_p \d\tilde{A}\\
F_4&=0
~.
}}
Once again we see that, contrary to the D3 case, the spacetime is neither flat nor empty in the spatial infinity limit: rather it is conformal to $\mathbb{R}^{1,2}\times C(\tilde{M}_6)$, where the latter factor is the metric cone over $\tilde{M}_6$.


Upon dimensional reduction on $\psi$ the M2 branes become D2 along $\mathbb{R}^{1,2}$, whose transverse space would be the cone over $\tilde{M}_6$. As 
in section \ref{sec:5} we find,
\begin{equation}
\star F_4 = -\frac{3\hat{Q}}{32 l_p}\sqrt{w(\theta)} v_6~.
\end{equation}
Since the cycle $\tilde{M}_6$ collapses in the transverse space at $r=0$,
\begin{equation}
\d\star F_4 = -\frac{3\hat{Q}}{32 l_p}\sqrt{w(\theta)} \delta(r) \d r\wedge v_6~.
\end{equation}
We compute the quantized D2 charge and obtain a similar result to the supersymmetric case, up to a factor of 2 difference arising from the different ranges of $\alpha$ and $\psi$, 
\bea
Q_{D2}=\frac{27\hat{Q}}{8192\pi^5l_p\alpha'^{5/2}}\int_0^{2\pi}d\alpha \int_0^{\pi}d\theta \frac{\sin\theta}{a(\theta)^{3/2}} \int_{S^3}d\Omega_3\int_0^{\frac{\pi}{2}}d\mu\sin^3\mu\cos\mu,\nn 
\eea
leading to a relation between the constant in the harmonic function and the number of D2 branes,
\bea
\hat{Q}=\frac{2048}{27}\pi^2l_p\alpha'^{5/2}N_{D2}~.
\eea
Since $F_2$ is not related to the D2 brane charge, we will only consider it as a geometric flux. For the same reasons as in the supersymmetric case, cf.~section \ref{sec:5}, 
there are no D4/D6 branes.

\subsection{The NATD}\label{sec:41}

We can now take the NAT dual of the background \eqref{nons}. The NS-NS sector of the resulting background is given by,
\bea \label{MasslessNSN0}
\hat{ds}^2&=& \frac{r}{2l_p}H^{-1/2}ds^2(\mathbb{R}^{1,2})+\Lambda^2\big(\frac{4}{r^2}dr^2+ 3 d\mu^2+\frac{1}{(1+\cos^2\theta)^2}d\theta^2\nn \\&~& +4 w(\theta) \sin^2\theta\d\alpha^2\big)+\frac{3\alpha'^2\Xi}{4M} [d(\rho\sin\chi)]^2\nn \\&~&+\frac{81}{256\alpha'\Delta}\bigg[\rho^2\Xi^2\cos^2\mu\sin^2\chi(d\xi)^2+\frac{1}{M}\bigg(\alpha'^2\rho^2\cos\chi d\rho+\Xi^2 d(\rho\cos\chi)\bigg)^2\bigg]  \nn \\
\hat{B}_2&=&\frac{81 \rho^2  \sin\chi\Xi}{256 \Delta}d\xi\wedge d\rho\chi  \nn \\
e^{-2\hat{\phi}}&=& e^{-2\phi}\Delta,\quad \Delta=\frac{27\Xi}{64\alpha'^3}\big[\cos^2\mu \Xi^2 + \alpha'^2\rho^2 K \big], \quad \Xi=\sin^2\mu \Lambda^2~, \quad \Lambda=\frac{1}{2}e^{\phi/3}rH^{1/6}~,
\eea
where we have defined the following one-form,
\bea
d\rho\chi&=&\Big(\rho K d\chi -\cos\chi \sin\chi\sin^2\mu d\rho\Big)~,
\eea
and included the following definitions,
\bea \label{QKMb}
K&=&\cos^2\mu\cos^2\chi+\sin^2\chi,\nn \\
M&=&\alpha'^2\rho^2\cos^2\chi+\Xi^2~.
\eea
The RR sector is given by
\bea
\hat{F}_1&=&-\frac{9l_p\sqrt{\Xi}}{16\sqrt{\alpha'}\Lambda^2}\bigg[\sqrt{\Xi} d(\rho\cos\chi)  +2\rho\cos\mu\cos\chi \Lambda d\mu\bigg],\nn \\
\hat{F}_3&=& \bigg(\frac{9l_p\sqrt{\alpha'}\rho\sin\theta}{16a(\theta)^2\sqrt{w(\theta)}}d\rho+\frac{18l_p^2\sqrt{w(\theta)}\Lambda^3\Xi^{3/2}H'}{\alpha'^{3/2}r^2a(\theta)H^{3/2}}\cos\mu\sin\theta d\mu \bigg)\wedge d\theta\wedge d\alpha \nn \\&~& 
+\frac{729l_p\rho \Xi^{3/2}\cos\mu\sin\chi}{4096\alpha'^{5/2}\Lambda^2 \Delta}\bigg(-2\cos^2\mu \Lambda\Xi^2 d\mu\wedge d(\rho\sin\chi) \nn \\&~&+\alpha'^2\rho^2 \Big[2\Lambda \sin\chi d\mu +\cos\mu\cos\chi\sqrt{\Xi} d\chi \Big]\wedge d\rho\bigg)\wedge d\xi,
\nn \\
\hat{F}_5&=&\frac{9}{32\alpha'^{3/2}}v_4\wedge\bigg(8\frac{l_p^2\alpha'^2\rho H'}{ r^2 H^{3/2}} d\rho +\frac{l_p\cos\mu\sin\theta\Xi^{3/2}}{a(\theta)w(\theta)\Lambda\sin\theta} d\mu \bigg)
\nn \\&~&-\frac{729l_p\rho^2\cos\mu\sin\chi\Xi^{5/2}}{4096r^2\alpha'^{3/2}a(\theta)^2\sqrt{w(\theta)}H^{3/2}\Delta}d\alpha\wedge d\theta\wedge d\xi \wedge\nn \\&~& \bigg(r^2\cos\mu\sin\theta\sqrt{\Xi} H^{3/2} d\rho\wedge d\chi  +32l_pa(\theta) w(\theta)\sin\theta H'\Lambda^3 d\mu\wedge d\rho\chi  \bigg)~,
 \label{MasslessRRN0}
\eea
where $a(\theta)=2(1+\cos^2\theta)$ and
$v_4=-\frac{r^2}{4l_p^2\sqrt{H}}dr\wedge dx_0\wedge dx_1\wedge dx_2$.

\subsubsection{Brane configuration and charges}\label{sec:42}

As in the supersymmetric reduction, spacetime is singular at $\mu=0$, which corresponds to the fixed locus of the $SU(2)$ isometry before duality. We thus first compute the NS5 charge by integrating $H_3$ on the cycle $(\Sigma_3[\rho,\chi,\xi], \mu=0)$, on which $H_3$ simplifies to,
\bea
H_3&=&\frac{3}{4}\alpha' \sin\chi d\xi\wedge d\chi\wedge d\rho~,
\eea
so that,
\bea\label{QNS5D2N0}
Q_{NS5}=\frac{1}{2\kappa_{10}^2T_{NS5}}\frac{3\alpha'}{4}\int_0^{\rho_0}d\rho\int_0^{\pi}\sin\chi d\chi\int_0^{2\pi}d\xi=\frac{3\rho_0}{4\pi}=N_{NS5}~.
\eea
For the charge to be quantized we need $\rho_0=\frac{4n\pi}{3}$. This is compatible with the condition (\ref{botrick}) which leads to $\rho\in [\frac{4(n-1)\pi}{3},\frac{4n\pi}{3}]$ and a large gauge transformation on $B_2$.
We can now examine the metric close to $\mu=0$, with $\nu=\mu^2$,
\bea\label{102er}
ds^2_{\mu\to 0}&=&\frac{r}{2l_p\sqrt{H(r)}}\Big[ds^2(\mathbb{R}^{1,2})+H(r) \big(dr^2+ r^2\big(\frac{1}{a(\theta)^2}d\theta^2 + \frac{1}{4a(\theta)}\sin^2\theta d\alpha^2 \big)\big)\Big]\nn\\
&+&\frac{1}{\nu}\Big[\frac{3}{32l_pr^3\sqrt{H(r)}}\bigg(16 l_p^2\alpha'^2 d\rho^2+r^6H(r)\Big[d\nu^2+\nu^2\big(d\chi^2 +\sin^2\chi d\xi^2\big)\Big]\bigg)\Big]
~,\eea
where $\nu^{-1}$ is the harmonic function for NS5 branes along the $(\mathbb{R}^{1,2},r,\theta,\alpha)$ directions. As in the previous example, the NS5 branes are located at the singularity $\mu=0$ and are smeared along the $\rho$ direction.

The Page forms are given by
\bea
\tilde{F}_3&=&\frac{9l_p\sqrt{\alpha'}\rho}{64a(\theta)^2\sqrt{w(\theta)}}\big[-4\sin\theta d\alpha\wedge d\theta\wedge d\rho\nn \\&~&+6a(\theta)^2\sqrt{w(\theta)}\cos\mu\sin\mu \sin\chi d\mu\wedge d(\rho\sin\chi)\big]\nn \\&~&
-\frac{27\hat{Q}}{256 l_p\alpha'^{3/2}a(\theta)}\cos\mu\sin^2\mu\sin\theta d\alpha\wedge d\theta\wedge d\mu\nn \\
\tilde{F}_5&=&-\frac{27l_p^2\sqrt{\alpha'}\hat{Q}}{2r^9H^{3/2}}\rho d\rho\wedge v_4+\frac{9r^3\sqrt{H}}{64\alpha'^{3/2}}\cos\mu\sin^3\mu d\mu\wedge v_4\nn \\&~& -\frac{27l_p\alpha'^{3/2}}{64a(\theta)^2\sqrt{w(\theta)}}\rho^2\sin\theta\sin\chi d\alpha\wedge d\theta\wedge d\xi\wedge d\rho \wedge d\chi~, 
%
\eea
with $\tilde{F}_1=\hat{F}_1$ given in \eqref{MasslessRRN0}. We will focus on the components that are proportional to $\hat{Q}$, whereas the remaining term will only be considered as geometric flux.
Integrating the $(\alpha,\theta,\mu)$ term in $\tilde{F}_3$ we obtain, 
\bea
Q_{D5}=\frac{1}{2\kappa_{10}^2T_{D5}}\frac{27\hat{Q}}{256 l_p\alpha'^{3/2}}\int_0^{2\pi}d\alpha\int_0^{\pi}\frac{\sin\theta}{a(\theta)^2}d\theta\int_0^{\frac{\pi}{2}}\cos\mu\sin^3\mu d\mu=N_{D5}~,
\eea
leading to a relation between the constant in the harmonic function and the number of D5 branes,
\bea
\hat{Q}=\frac{4096}{27}\pi l_p\alpha'^{5/2}N_{D5}~.
\eea
If we further consider the change in the Page forms under a large gauge transformation in $B_2$\footnote{The large gauge transformation has the same expression each time: see (\ref{deltaB}) or (\ref{deltaB2})}, it is the D3 charge which is created and we find $\Delta Q_{D3}=n N_{D5}$.

In the same spirit as before, we would then have the following brane configuration:
\begin{center}
	\begin{tabular}{ |c|ccc|ccccccc|} 
		\hline
		& 0&1&2&$ r$&$\mu$&$\theta$&$\alpha$ &$\rho$&$\chi$&$\xi$ \\
		\hline
		$NS5$ & $\times$ &$\times$ &$\times$ &$\times$ & & $\times$ &$\times$ & & & \\ 
		$D5$  & $\times$&$\times$&$\times$& & & & &$\times$ &$\times$ & $\times$\\ 
		$D3$ & $\times$&$\times$& $\times$ & &  & & &$\times$ & & \\ 
		\hline
	\end{tabular}
\end{center}


\subsubsection{The spatial infinity limit}\label{sec:44}


The NS-NS sector of the spatial infinity limit of the non-supersymmetric D2-brane NATD solution is obtained by  setting  $H(r)= 1$ in ~\eqref{MasslessNSN0}-\eqref{MasslessRRN0}, 
\bea \label{MasslessNSAsym}
\hat{ds}^2&=& \frac{r}{2l_p}\Big(ds^2(\mathbb{R}^{1,2})+dr^2+r^2\big( 3 d\mu^2+\frac{1}{(1+\cos^2\theta)^2}d\theta^2 +4 w(\theta) \sin^2\theta\d\alpha^2\big)\Big)\nn \\&~&+\frac{6l_p\alpha'^2\Xi}{M} [d(\rho\sin\chi)]^2+\frac{81}{16384l_p^2\alpha'\Delta}\bigg[\rho^2\Xi^2\cos^2\mu\sin^2\chi(d\xi)^2\nn \\&~&+\frac{1}{M}\bigg(64l_p^2\alpha'^2\rho^2\cos\chi d\rho+\Xi^2 d(\rho\cos\chi)\bigg)^2\bigg] \nn \\
\hat{B}_2&=&\frac{81 \rho^2  \sin\chi\Xi}{2048l_p \Delta}d\xi\wedge d\rho\chi  \nn \\
e^{-2\hat{\phi}}&=& e^{-2\phi}\Delta,\quad \Delta=\frac{27\Xi}{32768l_p^3\alpha'^3}\big[\cos^2\mu \Xi^2 + 64l_p^2 \alpha'^2\rho^2 K \big], \quad \Xi=r^3 \sin^2\mu ~,
\eea
where we have defined the following one-form,
\bea
d\rho\chi&=&\Big(\rho K d\chi -\cos\chi \sin\chi\sin^2\mu d\rho\Big)~,
\eea
and included the following definitions,
\bea \label{QKMc}
K&=&\cos^2\mu\cos^2\chi+\sin^2\chi,\nn \\
M&=&64l_p^2\alpha'^2\rho^2\cos^2\chi+\Xi^2~.
\eea
The RR sector is given by,
\bea \label{MasslessRRAsym}
\hat{F}_1&=&-\frac{9l_p\sqrt{\Xi}}{16r^3\sqrt{\alpha'}}\bigg[\sqrt{\Xi} d(\rho\cos\chi)  +2r^{3/2}\rho\cos\mu\cos\chi d\mu\bigg],\nn \\
\hat{F}_3&=& \frac{9l_p\sqrt{\alpha'}\rho\sin\theta}{16a(\theta)^2\sqrt{w(\theta)}}d\theta\wedge d\alpha\wedge d\rho 
\nn \\&~& +\frac{729\rho \Xi^{3/2}\cos\mu\sin\chi}{1048576r^3l_p^2\alpha'^{5/2} \Delta}\bigg(r^{3/2}\cos^2\mu \Xi^2 d\mu\wedge d\xi\wedge d(\rho\sin\chi) \nn \\&~&+32l_p^2\alpha'^2\rho^2 \Big[2r^{3/2}\sin\chi d\mu +\cos\mu\cos\chi\sqrt{\Xi} d\chi \Big]\wedge d\mu\wedge d\xi\bigg), \\
\hat{F}_5&=&\frac{9\cos\mu\sin\theta\Xi^{3/2}}{256r^{3/2}\alpha'^{3/2}a(\theta)w(\theta)\sin\theta}v_4\wedge d\mu 
\nn\\&~&+\frac{729\rho^2\cos^2\mu\sin\chi\sin\theta\Xi^3}{2097152l_p^2\alpha'^{3/2}a(\theta)^2\sqrt{w(\theta)}\Delta}d\theta\wedge d\alpha\wedge d\xi \wedge  d\rho\wedge d\chi \nn ,
\eea
with $v_4=-\frac{r^2}{4l_p^2}dr\wedge dx_0\wedge dx_1\wedge dx_2$.

The ten-dimensional  spacetime is 
a foliation over the $r$-coordinate with leaves of the form 
of a warped product $\mathbb{R}^{1,2}\times\tilde{M}_6$. The general structure of the leaves is very similar to that 
of section \ref{sec:54}, and can be analyzed in the same way: at fixed $r$, the space $\tilde{M}_6$ can be thought of as a fibration of the space $\tilde{N}_3$ parameterized by $(\rho,\chi,\xi)$ fibered over the base $\tilde{M}_3$ parameterized 
by $(\mu,\theta,\alpha)$. The topology of  $\tilde{M}_3$    is that of an $S^2$ parameterized by $(\theta, \alpha)$ times the interval parameterized by $\mu$. 

The range of the coordinate $\rho$ was constrained by flux quantization to be the interval specified in section \ref{sec:42}. Moreover, over a fixed base point 
$(\mu,\theta,\alpha)\in\tilde{M}_3$, the coordinates   $(\chi,\xi)$ parameterize a smooth $S^2$ 
provided we take $\xi\in[0,2\pi]$, $\chi\in[0,\pi]$. This can already be seen from the geometry near the location of the NS5 branes, cf.~\eqref{102er}. More generally 
the geometry of the  $\tilde{N}_3$ fiber 
over a fixed point in $\tilde{M}_3$ is  a smooth $S^2$   parameterized by $(\chi,\xi)$ fibered over 
the interval parameterized by $\rho$.

As in the supersymmetric D2 case, we have thus 
been able to specify the ranges of all coordinates parameterizing the NATD space. 
Once this result has been established for the leaf of the $r$-foliation at spatial infinity, it remains 
valid for finite $r$ and applies also to the full interpolating solution \eqref{MasslessNSN0}. 
The near-horizon limit is obtained by substituting $H\rightarrow\tfrac{\hat{Q}}{r^6}$ in \eqref{MasslessNSN0}, and results in an AdS$_4$ factor exactly as is the supersymmetric case.

\section{D2 from reduction on $S^7$}\label{sec:3}

Here we  consider the  reduction of the M2 brane background of ~(\ref{1}), to IIA along $\psi_1$,
\bea\label{S7red}
ds_{10}^2&=&\frac{r}{2l_p}\cos\frac{\mu}{2} \Big[H(r)^{-1/2}ds^2(\mathbb{R}^{1,2})+H(r)^{1/2}(\d r^2+\frac{1}{4}r^2(\sin^2\frac{\mu}{2}\Sigma_i^2+
\cos^2\frac{\mu}{2}ds^2(\Omega_2)+d\mu^2))\Big]\nonumber \\
B_2&=&0,\quad e^{2\Phi}=\frac{r^3}{8l_p^3}\sqrt{H(r)}\cos^3\frac{\mu}{2} \nonumber \\
 F_2&=&-l_pd\Omega_2,\quad F_4=-dH^{-1}\wedge d\text{vol}_3~,
\eea
with $\Omega_2$ representing an $S^2$ with coordinates ($\theta_1,
\phi_1$) leftover from the $\sigma_i$ in~(\ref{1}).
The near horizon limit of this solution and its NATD were given explicitly in \cite{Zayas:2015azn}.

We can see the presence of D2 branes from $\star F_4$,
\begin{equation}
\star F_4 = -\frac{3 \hat{Q}}{64 l_p} \cos\frac{\mu}{2} v_6~,
\end{equation}
where $v_6$ is the volume form of the 6-dimensional space $M_6$ (along $\mu,\d\Omega_2$ and $\d\Omega_3$). If we take the transverse space to be the cone over $M_6$, this cycle collapses at $r=0$, where we can see the D2 brane,
\begin{equation}
\d\star F_4 = -\frac{3 \hat{Q}}{64 l_p} \cos\frac{\mu}{2} \delta(r) \d r\wedge v_6~.
\end{equation}
Upon quantizing the flux, we obtain
\bea
Q_{D2}=\frac{1}{2\kappa_{10}^2T_{D2}}\int_{M_6} \star F_4 = -\frac{3\hat{Q}}{2048\pi^5l_p\alpha'^{5/2}}\int_{S^2}d\Omega_2 \int_{S^3}d\Omega_3\int_0^{\frac{\pi}{2}}d\mu\sin^3\frac{\mu}{2}\cos^3\frac{\mu}{2}
~,\nn 
\eea
leading to,
\bea\label{112ert}
\hat{Q}=128\pi^2l_p\alpha'^{5/2}N_{D2}~.
\eea
On the other hand, as is the case for the near-horizon limit, $F_2$  is sourced by a $D6$ brane along $\mathbb{R}^{(1,2)}, r, \Omega_3$ and located at $\mu={\pi}$, where the 2-sphere $\Omega_2$ collapses. 
As shown in \cite{Zayas:2015azn}, the metric in the vicinity of $\mu={\pi}$ is singular and takes the precise form of the metric near a D6 brane source.   
The charge is given by:
\begin{equation}\label{113ade}
Q_{D6} = \frac{1}{2\kappa_{10}^2T_{D6}} \int F_2=-\frac{2 l_p}{\sqrt{\alpha'}}~.
\end{equation}

The brane configuration is thus the following:

\begin{center}
	\begin{tabular}{ |c|ccc|ccccccc|} 
		\hline
		& 0&1&2&$ r$&$\mu$&$\theta_1$&$\phi_1$ &$\theta_2$&$\phi_2$&$\psi_2$ \\
		\hline
		$D2$ & $\times$ &$\times$ &$\times$ & & & & & & & \\ 
		$D6$  & $\times$&$\times$&$\times$&$\times$ & & & &$\times$ &$\times$ & $\times$\\ 
		\hline
	\end{tabular}
\end{center}

Note that the 3-sphere $\Omega_3$, on which we will now dualize, is transverse to the D2 but parallel to the D6. We will now see how both will behave under NATD.

\subsection{The NATD}\label{sec:31}

The background resulting from the application of NATD on the $\Sigma_i$ reads,
\bea
\label{NATDofRoundS7}
\hat{ds}^2&=&\frac{r\cos\frac{\mu}{2}}{2l_p}\bigg[H(r)^{-1/2}ds^2(\mathbb{R}^{1,2})+H(r)^{1/2}\Big(dr^2+r^2\big[d\mu^2+\cos^2\frac{\mu}{2}ds^2(\Omega_2)\nn \\&~&+\frac{9r^3\sqrt{H(r)}\rho^2\cos\frac{\mu}{2}\sin^4\frac{\mu}{2}}{4096l_p^2\alpha'\Delta}ds^2(d\chi^2+\sin^2\chi d\xi^2)\big]\Big)\bigg]+\frac{9l_p\alpha'^2}{8r^3\sqrt{H(r)}\cos\frac{\mu}{2}\sin^2\frac{\mu}{2}}d\rho^2, \nn \\
\hat{B}_2&=&\frac{27r^3\rho^3\cos\frac{\mu}{2}\sin^2\frac{\mu}{2}}{4096l_p\Delta}\sin\chi d\xi\wedge d\chi, \quad
e^{-2\hat{\Phi}}=\frac{8l_p^3\Delta}{r^3\sqrt{H(r)}\cos^3\frac{\mu}{2}},\nn \\
\Delta&=&\frac{r^3\sqrt{H(r)}\cos\frac{\mu}{2}\sin^2\frac{\mu}{2}}{512l_p^3\alpha'^3}(9l_p^2\alpha'^2\rho^2+r^6H(r)\cos^2\frac{\mu}{2}\sin^2\frac{\mu}{2})~,
\eea
and,
\bea \label{S7RR}
\hat{F}_3&=&-\frac{9}{64}l_p\sqrt{\alpha'}\rho d\Omega_2\wedge d\rho+\frac{3\hat{Q}}{64 l_p\alpha'^{3/2}}\cos^3\frac{\mu}{2}\sin^3\frac{\mu}{2} d\Omega_2\wedge d\mu \nn \\
\hat{F}_5&=&-\frac{27 \hat{Q} l_p^2\sqrt{\alpha'}}{8r^9H(r)^{3/2}\cos^2\frac{\mu}{2}}\rho d\rho\wedge v_4+\frac{r^3\sqrt{H}\sin^3\frac{\mu}{2}}{8\alpha'^{3/2}\cos\frac{\mu}{2}}d\mu \wedge v_4 \\ &~&
+\frac{27 r^3\sqrt{H}\rho^2\cos^3\frac{\mu}{2}\sin^5\frac{\mu}{2}}{524288l_p^2\alpha'^{3/2}\Delta}\Big(2r^6 H\sin\frac{\mu}{2}d\rho-6\hat{Q}\rho\cos\frac{\mu}{2} d\mu \Big)\wedge d\Omega_2\wedge \sin\chi d\chi\wedge d\xi~,\nn 
\eea
with $v_4=-\frac{r^2\cos^2\frac{\mu}{2}}{4l_p^2\sqrt{H}}dr\wedge dx_0\wedge dx_1\wedge dx_2$.

The Page five-form is given by,
\bea \label{PageF5}
\tilde{F}_5&=& -\frac{27 \hat{Q} l_p^2\sqrt{\alpha'}}{8r^9\cos^2\frac{\mu}{2}H^{3/2}}\rho v_4\wedge d\rho +\frac{r^3\sqrt{H}\sin^3\frac{\mu}{2}}{8\alpha'^{3/2}}v_4\wedge d\mu\nn \\&~&-\frac{27}{512}l_p\alpha'^{3/2}\rho^2\sin\chi d\Omega_2\wedge d\rho\wedge d\chi\wedge d\xi~.
\eea

\subsubsection{Brane configuration and charges}\label{sec:32}

We first compute the NS5 charge by integrating $H_3$ on the cycle $([\rho,\chi,\xi], \mu=0)$,
\bea
H_3&=&-\frac{3}{8}\alpha'\sin\chi d\rho\wedge d\chi \wedge d\xi\\
Q_{NS5} &=& \frac{1}{2\kappa_{10}^2T_{NS5}} \int H_3 = \frac{3\rho_{0}}{8\pi}~.\label{11a8}
\eea
$Q_{NS5}$ is quantized if  $\rho_0=L_{n}$, where we set $L_n:=\frac{8}{3}\pi n$. With $\rho \in [L_{n},L_{n+1}]$ and a suitable large gauge transformation on $B_2$, the relation \eqref{botrick} is satisfied.

The NS5 branes are also seen by zooming in on the singularity generated by the NATD at $\mu=0$,
\eq{\spl{
ds^2_{\mu\to 0}=\frac{r}{2l_p\sqrt{H(r)}}\Big[ds^2(\mathbb{R}^{1,2})&+H(r) \big(dr^2+\frac{r^2}{4} d\Omega_2^2\big)\Big]\\
&+\frac{1}{\nu}\Big[\frac{9l_p\alpha'^2}{2r^3\sqrt{H(r)}}d\rho^2+\frac{r^3\sqrt{H(r)}}{32l_p}\Big( d\nu^2+\nu^2d\tilde{\Omega}\Big) \Big]
~.}}
This is indeed consistent with the harmonic superposition rule, with harmonic function proportional to $\nu^{-1}$. This gives the characteristic NS5 brane configuration: along the $(\mathbb{R}^{1,2},r,\Omega_2)$ directions, located at $\mu=0$ and smeared along $\rho$.

Next we compute the quantized Page charges. We start with the dual of $F_4$ to track the D2. This corresponds to the terms proportional to $\hat{Q}$. Here only the $\tilde{F}_3=\hat{F}_3$  gives a non zero charge and we integrate the  ($\Omega_2,\mu$) term to find,
\bea\label{120reure}
Q_{D5}=\frac{\hat{Q}}{256l_p\pi\alpha'^{5/2}}=\frac12\pi  N_{D2}~,
\eea
where we took \eqref{112ert}  into account. We see that, as already noted in the near-horizon limit \cite{Zayas:2015azn},  $N_{D5}$  and $N_{D2}$ differ by a factor of $\tfrac{\pi}{2}$ and thus cannot both be integers. Indeed it is known that NATD generically maps integer charges  to non-integer ones \cite{Lozano:2014ata}.   
In the dual theory we are thus led to impose a different quantization condition: $\tfrac12\pi  N_{D2}\in\mathbb{Z}$, 
so that \eqref{120reure} is satisfied with $Q_{D5}\in\mathbb{Z}$. 
Moreover, we may perform a large gauge transformation on $B_2$ and find the resulting change in the Page charge for $F_5$, $\Delta Q_{D3}$,
\bea\label{122af}
\Delta Q_{D3}=\frac{n \hat{Q}}{256\pi l_p\alpha'^{5/2}}=n Q_{D5}~.
\eea
We can also track the D6 by looking at the dual of $F_2$, i.e.~the remaining components of the Page forms. These are found by integrating the terms not proportional to $\hat{Q}$ in \eqref{S7RR} and \eqref{PageF5}, which we label $ \tilde{F}_{3'}$ and $\tilde{F}_{5'}$.
We find,
\begin{eqnarray}
Q_{D5'} &=& \frac{1}{2\kappa_{10}^2T_{D5}}\int \tilde{F}_{3'}=-\frac{9 l_p (L_{n+1}^2-L_{n}^2)}{128\pi\sqrt{\alpha'}}=\frac14\pi (2n+1)  Q_{D6}\\
Q_{D3'} &=& \frac{1}{2\kappa_{10}^2T_{D3}}\int \tilde{F}_{5'}=-\frac{9 l_p (L_{n+1}^3-L_{n}^3)}{512\pi^2\sqrt{\alpha'}}
=\frac{1}{6}\pi  (3n^2+3n+1) Q_{D6}
~,
\end{eqnarray}
where in the last equalities on the right hand sides above we have taken \eqref{113ade} into account and the quantization of $\rho_{0}$ given below \eqref{11a8}. 
Similar to the case of $Q_{D5}$ above, we see  that $Q_{D5'}$, $Q_{D3'}$  cannot be integers if $Q_{D6}$ is integer. 
In the dual theory we are thus  led to impose a different quantization condition: $\tfrac{1}{12}\pi  Q_{D6}\in\mathbb{Z}$. 
Moreover, under a large gauge transformation of $B_2$, $Q_{D3'}$ is modified in the same fashion as $Q_{D3}$, cf.~\eqref{122af},
\bea
\Delta Q_{D3'}=\int -\Delta B_2\wedge \tilde{F}_3' &=& \frac{1}{2\kappa^2T_{D3}}\frac{9\pi l_p\alpha'^{3/2}}{64}\int_{L_n}^{L_{n+1}}\rho d\rho\int_0^{\pi}\sin\chi d\chi\int_0^{2\pi}d\xi\int d\Omega_2\nn \\
&=&-\frac{9 l_p n(L_{n+1}^2-L_{n}^2)}{128\pi\sqrt{\alpha'}} = n Q_{D5'}
\eea
The brane configuration is summarized in the following table.

\begin{center}
	\begin{tabular}{ |c|ccc|ccccccc|} 
		\hline
		& 0&1&2&$ r$&$\mu$&$\theta_1$&$\phi_1$ &$\rho$&$\chi$&$\xi$ \\
		\hline
		$NS5$ & $\times$ &$\times$ &$\times$ &$\times$ & &$\times$ &$\times$ & & & \\ 
		$D3$  & $\times$ &$\times$ &$\times$ & & & & &$\times$ & & \\ 
		$D5$  & $\times$ &$\times$ &$\times$ & & & & &$\times$ &$\times$ & $\times$\\ 
		$D3'$ & $\times$ &$\times$ &$\times$ &$\times$ & & & & & & \\ 
		$D5'$ & $\times$ &$\times$ &$\times$ &$\times$ & & & & &$\times$ & $\times$\\ 

		\hline
	\end{tabular}
\end{center}


\subsubsection{The spatial infinity limit}\label{sec:34}
The supergravity background corresponding to the NATD of the spatial infinity limit of (\ref{S7red}) is presented here,
\bea
\hat{ds}^2&=&\frac{r\cos\frac{\mu}{2}}{2l_p}\bigg[ds^2(\mathbb{R}^{1,2})+\Big(dr^2+r^2\big[d\mu^2+\cos^2\frac{\mu}{2}ds^2(\Omega_2)\nn \\&~&+\frac{9r^3\rho^2\cos\frac{\mu}{2}\sin^4\frac{\mu}{2}}{4096l_p^2\alpha'\Delta}ds^2(d\chi^2+\sin^2\chi d\xi^2)\big]\Big)\bigg]+\frac{9l_p\alpha'^2}{8r^3\cos\frac{\mu}{2}\sin^2\frac{\mu}{2}}d\rho^2, \nn \\
\hat{B}_2&=&\frac{27r^3\rho^3\cos\frac{\mu}{2}\sin^2\frac{\mu}{2}}{4096l_p\Delta}\sin\chi d\xi\wedge d\chi, \quad
e^{-2\hat{\Phi}}=\frac{8l_p^3\Delta}{r^3\cos^3\frac{\mu}{2}},\nn \\
\Delta&=&\frac{r^3\cos\frac{\mu}{2}\sin^2\frac{\mu}{2}}{512l_p^3\alpha'^3}(9l_p^2\alpha'^2\rho^2+r^6\cos^2\frac{\mu}{2}\sin^2\frac{\mu}{2})~,
\eea
and,
\bea
F_3&=&-\frac{9}{64}l_p\sqrt{\alpha'}\rho d\Omega_2\wedge d\rho \nn \\
F_5&=&\frac{r^3\sin^3\frac{\mu}{2}}{8\alpha'^{3/2}\cos\frac{\mu}{2}}d\mu \wedge v_4 
+\frac{27 r^9\rho^2\cos^3\frac{\mu}{2}\sin^6\frac{\mu}{2}}{262144l_p^2\alpha'^{3/2}\Delta}d\rho \wedge d\Omega_2\wedge \sin\chi d\chi\wedge d\xi~,\nn 
\eea
with $v_4=-\frac{r^2\cos^2\frac{\mu}{2}}{4l_p^2}dr\wedge dx_0\wedge dx_1\wedge dx_2$.  The surviving RR flux terms in the asymptotic limit ultimately arise from the charge created in the reduction of the parent M-theory background to Type IIA, and thus from the D6. The NS5 also survives since it comes from the singularity in the NATD.


\section{Domain wall supersymmetry equations}\label{sec:dw}

As already mentioned, (\ref{8}) is a supersymmetric domain wall (DW) solution in four-dimensional space, where the latter is viewed as a foliation,  parameterized by $r$, with $\mathbb{R}^{1,2}$ 
leaves. The supersymmetry conditions for $\mathcal{N}=1$ domain walls were written in \cite{Haack:2009jg} in generalized $G_2\times G_2$ form in 
eqs.~(2.5), (2.6) therein. For our purposes it would be more useful to recast these equations in terms of generalized pure spinors on $M_6$. Such a rewriting is indeed 
given in \cite{Haack:2009jg}, {\it cf.}~(2.7) therein. We will now review their results adapting them to our case.



The ansatz for the splitting of the metric and the flux is given by:
\begin{equation}\label{7split}
\begin{array}{rcl}
\d s^2 &=& e^{2 A} \d s^2(\mathbb{R}^{1,2}) + \d s^2(M_7)\\
F_t     &=& F + v_3 \wedge \star \lambda F~,
\end{array}
\end{equation}
where the warp factor $A$ and the dilaton $\phi$ are not constrained  at this point; $\lambda$ is an involution reversing 
the order of wedge products. The NS-NS form 
$H$ is assumed to be internal, i.e. to only have legs along $M_7$, and likewise for the internal RR flux $F$. The total flux $F_t$ is then chosen to be self-dual: for $F$ internal we get $\star\lambda F = v_3\wedge\star_7\lambda F$ and in ten Lorentzian dimensions $(\star\lambda)^2=1$. Unbroken supersymmetry of the solution implies on $M_7$ the existence of two Majorana spinors $\chi_1,\chi_2$ normalized so that $\chi_a^\dagger\chi_a=1$. This leads us to define a bispinor $\Psi$, which can also be viewed as a polyform via the Clifford map:
\begin{equation}\label{psi7}
\Psi = 8 \chi_1\otimes\chi_2^\dagger = \Psi_+ +i\,\Psi_-
~,
\end{equation}
where $\Psi_+$ and $\Psi_-$ are respectively the real-even and imaginary-odd parts of $\Psi$. We should be careful however about how the identification is imposed: odd dimensional Fierzing does not provide an isomorphism between bispinors and polyforms because the Clifford representation is not faithful, as can be confirmed by a simple count of dimensions. We thus need to choose the range of our identification. Here we take $\Psi$ to be self-dual as a polyform: $-i\star_7\lambda\Psi=\Psi$. This also means that the decomposition (\ref{psi7}) is only valid in the polyform space and that $\Psi_+$, $\Psi_-$ are not independent:
\eq{
\Psi_+=\star_7\lambda\Psi_-~.
}
These choices lead to the normalization:
\begin{equation}
\langle \Psi_+,\Psi_-\rangle = \frac{i}{2}\langle\Psi,\bar{\Psi}\rangle=8v_7~.
\end{equation}
We now have all the necessary ingredients to write the supersymmetry for IIA in terms of generalized spinors:
\begin{equation}\label{susy7}
\begin{array}{rcl}
\d{}_H(e^{3A-\phi}\Psi_+) &=& -e^{3A}\star_7\lambda F\\
\d{}_H(e^{2A-\phi}\Psi_-) &=& 0\\
\langle \Psi_-,F\rangle    &=& 0~.
\end{array}
\end{equation}
In order to match (\ref{4}), \eqref{5} we need to further split $M_7$ to $M_6$ plus a transverse direction parameterized by the coordinate $r$. The metric and fluxes thus decompose as follows:
\begin{equation}\label{split6}
\begin{array}{rcl}
\d{} s^2 &=& e^{2 Z}(e^{2a} \d{} s^2(\mathbb{R}^{1,2})+\d{} r^2) + \d{} s^2(M_6)\\
F       &=& F_i + \d{} r\wedge F_r\\
H       &=& H_i + \d{} r\wedge H_r~,
\end{array}
\end{equation}
where $a$ depends only on $r$, and $F_i,F_r,H_i,H_r$ only have legs on $M_6$. Note also that the expression $\d{} s^2(M_6)$ can depend on $r$, since it can include a warp factor for instance. The same split must then be performed for the spinors, by expressing  7D spinors in terms of 6D chiral spinors. Since we are splitting along $r$, $\gamma_r$ (in flat basis) becomes the chirality matrix for spinors of $M_6$. Thus we take:
\begin{equation}
\begin{array}{rclcrcl}
\eta_1 &:=& \sqrt{2}\, P_+ \chi_1       &,& \eta _2 &:=& \sqrt{2}\, P_-\chi_2 \\
\chi_1 &= & \dfrac{1}{\sqrt{2}} (\eta_1+\eta_1^c) &,& \chi_2  &= & \dfrac{1}{\sqrt{2}}(\eta_2+\eta_2^c)~,
\end{array}
\end{equation}
where $P_\pm := \frac{1}{2}(1 \pm \gamma_r)$. Introducing the following bispinors on $M_6$ (which can be viewed  equivalently,  
via 6D Fierzing and the Clifford map,  
as  polyforms or generalized spinors):
\begin{equation}\label{poly6}
\Phi_1 := 8e^{3Z-\phi}\eta_1\otimes \eta_2^\dagger \quad ,\quad \Phi_2 := 8e^{3Z-\phi} \eta_1\otimes \tilde{\eta}_2  ~,
\end{equation}
we get:
\begin{equation}\label{7to6}
\Psi_+ = e^{-3Z+\phi} ( \Re\Phi_2 + e^Z\d{} r\wedge \Re \Phi_1 )     \quad , \quad  \Psi_- = e^{-3Z+\phi} (\Im\Phi_1+e^Z\d{} r\wedge\Im\Phi_2) ~.
\end{equation}
The factor $e^{3Z-\phi}$ is  introduced here for future convenience; it is simply another choice of  normalization:
\begin{equation}\label{normphi}
i\langle \Phi_1,\bar{\Phi}_1 \rangle = i\langle \Phi_2,\bar{\Phi}_2\rangle =8 e^{6Z-2\phi} v_6 ~.
\end{equation}
We then substitute  (\ref{split6}) and (\ref{7to6}) into (\ref{susy7}), and decompose along $\d{} r$. We look for an expression solely in terms of polyforms on the internal space $M_6$, where $r$ is now considered as an external parameter:
\begin{equation}\label{susy6}
\begin{array}{rcl}
\d{}^H \, e^Z \Re \Phi_1 &=& e^{4Z} \star\lambda F_i + e^{-3a} \partial_r^H e^{3a} \Re\Phi_2\\
\d{}^H \Re \Phi_2        &=& -e^{2Z} \star \lambda F_r\\
\d{}^H e^{-Z} \Im\Phi_1  &=& 0 \\
\d{}^H \Im\Phi_2        &=& e^{-2a}\partial_r^H e^{2a-Z} \Im\Phi_1\\
\langle \Im \Phi_1,F_r \rangle +e^Z\langle \Im\Phi_2,F_i\rangle &=& 0~,
\end{array}
\end{equation}
where now $\d{}^H=\d{} + H_i\wedge$, $\partial_r^H = \partial_r +H_r\wedge$, and $\d{}$ acts only on the coordinates of $M_6$. Note also that $\langle,\rangle$ now refers to the 6D Mukai pairing.

\subsection{Supersymmetric D2}\label{sec:susydw}

We now want to check explicitly that the solution in (\ref{4}), \eqref{5} is compatible with the equations (\ref{susy6}). This amounts to defining two polyforms $\Phi_1,\Phi_2$, whose $SU(3)\times SU(3)$-structure carries the 6D part of the metric (\ref{4}), and which is solution of (\ref{susy6}). First we need to identify the various fields. Comparing (\ref{4}) and (\ref{split6}) we find,
\begin{equation}
e^a = \frac{1}{\sqrt{H}} \quad \mathrm{and} \quad e^Z = e^{\phi/3}H^{1/6}~.
\end{equation}
Since the fluxes are not given in the same formalism, we need to retrieve $F_6$ and $F_8$ from $F_2$ and $F_4$ by Hodge duality, in order to build the total flux polyform $F_t$ in the democratic formalism. If we write $F_{nd}=F_2+F_4$, the total flux of the solution (\ref{5}) we find,
\[F_t = F_{nd} + \star_{10}\lambda F_{nd}~.\]
 This leads to,
\begin{equation}
\begin{array}{rcl}
F_i &=& F_2 +\star_{10}\lambda F_4 = \d{} f(\d{}\psi+A) + \frac{H'}{\sqrt{H}} e^{-4Z} v_6 \\
F_r &=& 0 \\
H_i &=& 0 \\
H_r &=& 0~,
\end{array}
\end{equation}
where $v_6$ is the volume form of $M_6$ taking into account the warp factor.
We can now use the results from section \ref{GCG} to define our polyforms $\Phi_1$ and $\Phi_2$. Our ansatz will introduce several functions of $\theta$ as supplementary degrees of freedom that should  enable us to find a solution of the DW equations.

We begin with the local $SU(2)$-structure, given by the K\"{a}hler structure  of ${B}_4$. We denote by $\hat{j}$ the K\"{a}hler form and $\hat{\omega}$ a holomorphic 2-form  normalized so that,
\begin{equation}\label{su2}
\begin{array}{rcl}
\hat{j}\wedge\hat{\omega} &=& \hat{\omega}\wedge\hat{\omega} = 0 \\
\hat{\omega}\wedge\hat{\omega}^* &=& 2 \hat{j}\wedge\hat{j}~.
\end{array}
\end{equation}
Note that 
$\hat{j}$ is global but $\hat{\omega}$ can only be defined locally. 
Furthermore 
we  define,
\begin{equation}
\begin{array}{rcl}
\tilde{\omega} &=& e^{2i(\psi+\zeta)} \hat{\omega}\\
j              &=& \frac{1}{4}r^2e^{2Z} \left( \cos\theta \hat{j}+\sin\theta \Re\tilde{\omega}  \right)\\
\omega         &=& \frac{1}{4}r^2e^{2Z+2i\alpha} \left( \cos\theta \Re \tilde{\omega}-\sin\theta \hat{j} + i\Im\tilde{\omega} \right)\\
K              &=& \frac{1}{2}re^{Z+i\beta} \left(\frac{1}{1+\cos^2\theta}\d{} \theta + i\sin\theta(\d{}\psi+A)   \right)~.
\end{array}
\end{equation}
Finally the polyforms (or, equivalently, the generalized spinors) defining the $SU(3)\times SU(3)$-structure are given by,
\begin{equation}
\begin{array}{rcl}
\Phi_1 &=& \sqrt{H}\, \bar{K}\wedge\left(e^{i\nu}\cos\varphi\,\bar{\omega}-\sin\varphi\, e^{i\,j}\right)\\
\Phi_2 &=& \sqrt{H}\, e^{-\frac{1}{2}K\wedge\bar{K}}\, \left(e^{-i\nu}\cos\varphi \,e^{i\, j}+\sin\varphi\,\bar{\omega}\right)~,
\end{array}
\end{equation}
where the factor $\sqrt{H}$ has been added to match the normalization (\ref{normphi}),
\begin{equation}
i\langle \Phi_1,\bar{\Phi}_1 \rangle = i\langle \Phi_2,\bar{\Phi}_2\rangle =8 e^{6Z-2\phi} v_6 = 8 H\, v_6~.
\end{equation}
We are thus left with five undetermined functions of $\theta$ ($\alpha,\beta,\zeta,\nu$ and $\varphi$) that should provide enough freedom for a solution of the DW supersymmetry equations: $\alpha$ and $\zeta$ act as rotation of the local $SU(2)$ and, since the $SU(2)$-structures span a two-sphere, they can be respectively seen as the intrinsic rotation and precession; $\beta$ is merely a modification of the phase of the vielbein one-form $K$; the meaning of $\nu$ and $\varphi$ is explained in section \ref{GCG},  recall in particular that $\varphi$ must vanish at $\theta=0,\pi$. Note also that a global phase of $\Phi_2$ can be absorbed in $\nu$ and $\alpha$ whereas a global phase 
of $\Phi_1$ can be absorbed in $\beta$.

{\it Solution}

Note first that in the near horizon limit, the $SU(3)\times SU(3)$ is in fact pure $SU(3)$, {i.e.}~$\varphi=0$. Thus if our ansatz is correct ($\varphi$ is function of $\theta$ only), $\varphi$ should remain constant to match the near horizon limit. Looking at the first equation of (\ref{susy6}), the scalar term gives straightforwardly,
\[\cos\varphi\cos\nu = 1~.\]
This is consistent with the ansatz, and also gives  information about $\nu$. We get
\[\varphi=0 \quad,\quad \nu =0~.\]
The structure is then pure $SU(3)$ all along the $r$ coordinate. Moreover  $\alpha$ and $\beta$ now play the same role: a global phase shift of the holomorphic 3-form. $\beta$ can thus be absorbed by a redefinition of $\alpha$, and be set to 0. At this point, the second and fifth equation of (\ref{susy6}) are satisfied. 
Moreover the three-form part of the fourth equation of (\ref{susy6})  leads to $2\alpha=-\frac{\pi}{2}$. All the remaining terms are then proportional to $\zeta'$ so that $\zeta$ has to be constant. Taking $\zeta=0$ then solves (\ref{susy6}).


\subsection{Mass Deformation}\label{sec:susymd}

The only difference between the full interpolating brane solution and its AdS$_4$ near horizon limit 
is a modification of the function $H(r)$. The background (\ref{4}), \eqref{5} is a genuine solution under the sole condition that $H(r)$ is harmonic in the transverse space $\mathbb{R}^7$. The interpolating solution corresponds to the most general choice of $H(r)$, whereas the near horizon limit and the spatial infinity limit correspond respectively to the choices $H(r)=Q/r^6$ and $H(r)=1$.
We would then expect that finding a massive deformation of the interpolating solution would amount to adding the correct $r$-dependence in the different functions of the massive deformation.

\paragraph{Consequences of the massive deformation:}
For the massive deformation of the near horizon limit, the $SU(3)\times SU(3)$ structure is no longer pure $SU(3)$ \cite{Petrini:2009ur,Lust:2009mb}, and this will obviously be the case for the interpolating solution. It will now be necessary to switch on the function $\nu$, $\varphi$ and $\zeta$, leading to our first source of complication. It is also important to notice that the base and fiber of $M_6$ get a different warp factor, and our ansatz should take that into account. We also expect the massive deformation to switch on all fluxes. Switching on the three-form $H$ also impacts   equation (\ref{su2}) by twisting the derivative.

Taking all the above into account, let us define the following ansatz for the $SU(3)\times SU(3)$ structure and the fluxes. 
We first define the local $SU(2)$ and one-form similarly to (\ref{rotsu2}),
\begin{equation}\label{rotsu2}
\begin{array}{rcl}
\tilde{\omega} &=& e^{2i(\psi+\zeta)} \hat{\omega}\\
j              &=& e^{2B} \left( \cos\gamma \hat{j}+\sin\gamma \Re\tilde{\omega}  \right)\\
\omega         &=& e^{2B+2i\alpha} \left( \cos\gamma \Re \tilde{\omega}-\sin\gamma \hat{j} + i\Im\tilde{\omega} \right)\\
K              &=& e^{C+i\beta} \left(f(\theta) \d{} \theta + i\sin\theta(\d{}\psi+A)   \right)~.
\end{array}
\end{equation}
The $SU(3)\times SU(3)$ structure is now given by,
\begin{equation}
\begin{array}{rcl}
\Phi_1 &=& e^{3Z-\phi}\, \bar{K}\wedge\left( e^{i\nu}\cos\varphi\,\bar{\omega}-\sin\varphi\, e^{i\,j}\right)\\
\Phi_2 &=& e^{3Z-\phi}\, e^{-\frac{1}{2}K\wedge\bar{K}}\, \left( e^{-i\nu}\cos\varphi \,e^{i\, j}+\sin\varphi\,\bar{\omega}\right)~.
\end{array}
\end{equation}
This corresponds to the ansatz of (\ref{split6}). The internal metric of $M_6$ follows from the $SU(3)\times SU(3)$,
\begin{equation}
\d s^2(M_6) = e^{2B} \d s^2({B}_4) + e^{2C} \left(f(\theta)^2 \d \theta^2 + \sin^2\theta (\d\psi+A)^2 \right)~.
\end{equation}
For The Bianchi Identities to be automatically satisfied we will rather define the fluxes by their potentials: the two-form $B$ and the odd polyform $C$,
\begin{equation}
\begin{array}{rcl}
B   &=& h \hat{j}   \\
C_1 &=& f_2 A \\   
C_3 &=& f_4 A\wedge \hat{j} \\ 
C_5 &=& f_6 A\wedge \hat{j}^2~.
\end{array}
\end{equation}
The fluxes are then, in the formulation of (\ref{split6}),
\begin{equation}
\begin{array}{rcl}
H_i &=& \d B \\
H_r &=& \partial_r B \\
F_i &=& \d^H C + m e^{-B} \\
F_r &=& \partial_r^H B~.
\end{array}
\end{equation}
All the warp factors $Z,B,C$, the dilaton $\phi$, the phases $\alpha,\beta,\gamma,\zeta,\varphi,\nu$ and the fluxes $h,f_2,f_4,f_6$ are allowed to depend on $r,\theta$.

Despite its generality, we have checked that this ansatz  does not  solve the supersymmetry equations (except for the solutions we already know, which are special cases thereof). 
Unfortunately, without further input,  relaxing the ansatz to allow for a dependence on more variables quickly becomes intractable.

 \section{Conclusions}\label{sec:concl}

 We have seen that having full-fledged  brane solutions, interpolating between the near-horizon and spatial infinity limit, may give a better handle on the brane configurations and 
 the global properties of the NATD. 
 In particular we have seen certain general features emerge. The NATD of the spatial infinity limit of 
  standard intersecting brane solutions is universal: it is given by a continuous linear distribution of NS5 branes along a half line with specific charge density. 
We have also provided additional examples where a general relation, observed previously in the NATD literature, between the Page charges generated by the NATD and their behavior under a large gauge transformation of the NS flux is obeyed. Since this behavior results from the non-trivial dependence of the Page charge on the $B_2$ field,  NATD naturally furnishes  several examples where the choice of $B_2$ plays an important role.

More generally we have seen that in cases where the brane configuration before NATD is not flat at spatial infinity, the NATD contains highly nontrivial RR fluxes even at spatial infinity. If the charges before NATD are related to the presence of branes, the latter can be tracked throughout the NATD. 
On the other hand,  the precise NS5-D(p+1)-D(p+3) brane
intersections underlying these solutions cannot be systematically 
identified with our approach. Indeed, we have not been able to describe these brane fluxes as resulting from backreaction (as dictated by the harmonic superposition rule) 
on some initial spacetime without branes. The exception to this statement is the case of the geometry near the locus of the NS5 branes. 
Let us also note that the spatial infinity limits of the NATD backgrounds presented here are highly nontrivial exact supergravity solutions in their own right, and they can be considered independently from the full interpolating intersecting brane solution.

In the case of the NATD of the D2 branes, proceeding by analogy to the NATD of the D3 brane, we have seen that cutting off the range of the $\rho$ coordinate at a finite value, in order to impose NS5 charge quantization, 
provides a prescription for assigning well-defined ranges to all dual coordinates. On the other hand, from a purely geometrical point of view this procedure renders the 
space geometrically incomplete. 
Ultimately such a procedure should be justified through a physical interpretation. In the case of the NATD of the D3 brane, 
such an interpretation was provided by the field theory dual proposed in \cite{Lozano:2016kum}, as reviewed in section \ref{sec:23}. 
It would be interesting to provide a similar interpretation for the NATD of the D2 branes of the present paper.

We have cast the supersymmetric D2 brane solution, arising from the reduction of M2 branes on seven-dimensional Sasaki-Einstein, in the language of generalized geometry 
pure spinor equations for domain walls. This framework allowed us to look for massive supersymmetric deformations of the D2 brane solutions, and we have been able to rule out 
a certain class of ans\"{a}tze. It would be interesting to try to construct these massive deformations explicitly, at least in a perturbative expansion in Romans mass as in \cite{Gaiotto:2009yz}. 
If they exist, these would be full interpolating intersecting brane solutions whose near-horizon limit coincides with the class of massive IIA AdS$_4\times M_6$ solutions of \cite{Lust:2009mb}.\footnote{This class includes the massive deformation of the IIA reduction of the eleven-dimensional AdS$_4\times M^{1,1,1}$ solution previously constructed in \cite{Petrini:2009ur}. Until recently these solutions were only known up to a system of two first-order nonlinear ordinary differential equations for two unknown functions. The class of analytic solutions of \cite{Passias:2018zlm} should include these (and other related solutions  such as \cite{Aharony:2010af,Tomasiello:2010zz,Guarino:2015jca}) as special cases. It would be interesting to construct the explicit dictionary between the two.}

It would also be interesting to cast the NATD of the supersymmetric solutions 
in the generalized geometry  formalism for domain walls, thus refining the general results of \cite{Itsios:2012dc,Kelekci:2014ima}. Besides providing a check of supersymmetry, this might give insight into the 
global structure of the solutions. In certain cases the duals might fall within the class recently examined in \cite{Macpherson:2017mvu}. 
We hope to return to these questions in the future.


\section*{Acknowledgments}
The authors would like to thank Leopoldo Pando Zayas for initial collaboration. RT and CAW would like to thank their respective institutions for hospitality. RT and CAW were supported in part by the National Research Foundation of South Africa 
and by the CNRS grant PICS 07552.  RT and CAW are especially grateful to Alan Cornell for providing travel support.
CAW was supported by the National Institute for Theoretical Physics and the Mandelstam Institute for Theoretical Physics in South Africa.


\appendix
\section{Conventions for $SU(3)$ and $SU(3)\times SU(3)$-structures}

In this section we review well-known results about  $SU(n)$-structures and generalized geometry. We follow here the approach presented in \cite{Koerber:2010bx} and \cite{Tsimpis:2016bbq}, but adapted to our conventions and notations.

\subsection{$SU(3)$-structures}

Let us first take an $SU(n)$-structure on a $2n$-dimensional manifold $M_{2n}$, given by a two-form $J$ and an $n$-form $\Omega$. In order to define our convention we introduce locally a vielbein $(e_k)$ and $z_k = e_{2k-1}+i\, e_{2k}$ such that $(J,\Omega)$ can be written:
\begin{eqnarray*}
J       &=& e_1\wedge e_2 +\cdots +e_{2n-1}\wedge e_{2n}=\frac{i}{2}(z_1\wedge\bar{z}_1+\cdots+z_n\wedge\bar{z}_n)\\
\Omega  &=& z_1\wedge \cdots \wedge z_n~.
\end{eqnarray*}
With this choice we find:
\begin{equation}\label{sun}
\begin{array}{rcl}
J \wedge\Omega    &=& 0\\
\Omega\wedge\bar{\Omega}  &=& (-1)^{\frac{1}{2}n(n+1)}\frac{(2i)^n}{n!} J^n~,
\end{array}
\end{equation}
which gives for our case $n=3$:
\begin{equation*}
\Omega\wedge\bar{\Omega}  = \frac{4}{3i} J^3~.
\end{equation*}
Note that $(J,\Omega)$ also define a metric $g$ and a volume form $v$ on $M_{2n}$ (and thus an orientation):
\begin{equation}\label{metric}
\begin{array}{rcl}
g &:=& e_1^2 +\cdots + e_n^2 \\
v &:=& \frac{1}{n!}J^n = e_1\wedge e_2 \wedge\cdots\wedge e_{2n}~.
\end{array}
\end{equation}
Now we can make the complex structure $I$ explicit, defined in such a way that $\Omega$ is $(n,0)$. For $n=3$:
\begin{eqnarray}
I_i^{\ j} &:=& J_{ik} g^{kj} \label{I}\\
I_i^{\ j} \Omega_{jkm} &=& i \Omega_{ikm} \label{IOmega}~.
\end{eqnarray}

\subsection{Pure Spinors}

Consider now a six-dimensional Riemannian spin manifold $M_6$, with metric $g$ and volume form $v_6$. Suppose that $M_6$ admits a chiral spinor $\gamma_7 \eta= \eta$, globally defined and nowhere-vanishing. $\eta$ is also pure and can be normalized to $\eta^\dagger\eta=1$. $\eta$ thus leads to an $SU(3)$-structure $(J,\Omega)$ by contraction with gamma matrices:
\begin{equation}\label{spin}
\begin{array}{rcl}
J_{ij}       &=& i\,\eta^\dagger \gamma_{ij} \eta=-i\,\tilde{\eta}\gamma_{ij} \eta^c\\
\Omega_{ijk} &=& \eta^\dagger \gamma_{ijk}\eta^c~.
\end{array}
\end{equation}
The choices here are made to be compatible with (\ref{sun}) and (\ref{metric}). This can be checked by Fierzing, using:
\begin{equation}\label{fierz}
\begin{array}{rcl}
\eta\eta^\dagger   &=& \frac{1}{4} (1+\frac{i}{2}J_{ij} \gamma^{ij})\,P^+\\
\eta\tilde{\eta}\, &=& \frac{1}{48} \bar{\Omega}_{ijk}\, P^+\gamma^{ijk}\\
\eta^c\eta^\dagger &=& -\frac{1}{48} \Omega_{ijk} \, P^-\gamma^{ijk}~,
\end{array}     
\end{equation}
where $ P^{\pm}$ is the projector to positive, negative chirality respectively.

\subsection{$SU(3)\times SU(3)$-structures}\label{GCG}

Suppose now that we have not only one but two spinors $\eta_1,\eta_2$ of positive chirality on $M_6$, normalized so that $\eta_1^\dagger\eta_1 = \eta_2^\dagger\eta_2=1$. Each defines its own $SU(3)$- and almost complex structure, according to (\ref{spin}). They both combine to form an $SU(3)\times SU(3)$-structure on $M_6$, given by two generalized spinors (i.e. spinors of $\mathrm{Spin}(3,3)$): 
\begin{equation}
\Psi_1 =8 \eta_1 \eta_2^\dagger  \quad , \quad \Psi_2 = 8\eta_1\tilde{\eta}_2~.
\end{equation}
Note that, thanks to Fierz isomorphism, generalized spinors can be seen equivalently as bispinors or polyforms (in even dimension). We thus want to express $\Psi_1,\Psi_2$ as polyforms. It is also important to notice that $\eta_1,\eta_2$ are not necessarily independent. Since $\eta_1$ is pure, the space of spinors can be constructed from $\eta_1$ or $\eta_1^c$ and (anti-)holomorphic gamma matrices. Note that we call a holomorphic gamma matrix the image of a holomorphic one-form in the Clifford algebra,  with respect to the complex structure defined by $\eta_1$. With this choice of conventions, the annihilators of $\eta_1$ are the anti-holomorphic gamma matrices, as can be seen by the following calculation. 
Consider a one-form $\bar{K}$ such that $\bar{K}\cdot\eta_1=\bar{K}_j\gamma^j\eta_1=0$, then:
\begin{eqnarray*}
I_{i}^{\, j} \bar{K}_j = J_{ij}\,\bar{K}^j &=& i \bar{K}^j\, \eta_1^\dagger \gamma_{ij} \eta_1 \\
                      &=& i \,\bar{K}^j \eta_1^\dagger \,(\gamma_i \gamma_j -g_{ij})\, \eta_1 \\
                      &=& -i\, \bar{K}_i~.
\end{eqnarray*}
It follows that any normalized spinor of positive chirality, such as $\eta_2$  in particular, can be written as:
\begin{equation}\label{eta2}
\eta_2 = e^{i\nu} \cos\varphi\, \eta_1+ \sin\varphi\,\chi~.
\end{equation}
Here $e^{i\nu} \cos \varphi = \eta_1^\dagger \eta_2$ and $\chi$ is a normalized spinor, orthogonal to $\eta_1$, and defined from an anti-holomorphic one-form $\bar{K}$ such that $\bar{K}\cdot K = 2$:
\eq{\chi = \frac{1}{2} \bar{K}_i\,\gamma^i\,\eta_1^c~.}
In this definition $K$ (and thus $\chi$) does not need to be globally well-defined, provided $\sin\varphi$ vanishes whenever the definition of $K$ fails. Moreover, $\chi$ (and $K$) defines locally another $SU(3)$-structure,  ``orthogonal'' to $\eta_1$'s:
\begin{equation}
\begin{array}{rcl}
J^\perp      &=& \chi^\dagger \gamma_{(2)} \chi = i K\wedge\bar{K}-J\\
\Omega^\perp &=& \chi^\dagger \gamma_{(3)} \chi^c = \frac{1}{2}K\cdot\bar{\Omega}\wedge K~.
\end{array}
\end{equation}
We can also define a local $SU(2)$-structure $(j,\omega)$:
\begin{equation}
\begin{array}{rcl}
j      &=& \frac{i}{2}(\eta_1^\dagger \gamma_{(2)}\eta_1 - \chi^\dagger\gamma_{(2)}\chi) = \frac{1}{2}(J-J^\perp)= J-\frac{i}{2}K\wedge\bar{K}\\
\omega &=& -\tilde{\chi}\gamma_{(2)} \eta_1^c = \frac{1}{2}\bar{K}\cdot\Omega~.
\end{array}
\end{equation}
The two orthogonal $SU(3)$-structures can be reconstructed back from the local $SU(2)$ and $K$: 
\begin{equation}
\begin{array}{rcl}
J            &=& j+\frac{i}{2}K\wedge\bar{K}\\
J^\perp      &=& -j+\frac{i}{2}K\wedge\bar{K}\\
\Omega       &=& \omega\wedge K\\
\Omega^\perp &=& \bar{\omega}\wedge K~.
\end{array}
\end{equation}
Now we can compute the generalized spinors using Fierz identities:
\begin{equation}\label{bispin}
\begin{array}{rcl}
\Psi_1 &=& e^{-\frac{1}{2}K\wedge\bar{K}}\, \left(e^{-i\nu}\cos\varphi \,e^{i\, j}+\sin\varphi\,\bar{\omega}\right)\\
\Psi_2 &=& \bar{K}\wedge\left(e^{i\nu}\cos\varphi\,\bar{\omega}-\sin\varphi\, e^{i\,j}\right)
\end{array}
\end{equation}
Finally we introduce the Mukai pairing of polyforms:
\begin{equation}
\langle \Phi,\Psi\rangle := (\Phi \wedge \lambda\Psi )|_{\mathrm{top}}~,
\end{equation}
where $\lambda$ swaps the order of the wedge product. This leads to the following 
normalization of the generalized spinors:
\begin{equation}
\begin{array}{rcl}
\langle \Psi_1,\bar{\Psi}_1\rangle=\langle \Psi_2,\bar{\Psi}_2\rangle= 8i\, v_6
\end{array}
\end{equation}

{\it Remarks on the conventions}

The conventions adopted here respect the general relation of $SU(n)$-structures (\ref{sun}) as well as the orientation (\ref{metric}). However this implies that the annihilators of a pure spinor of positive chirality are anti-holomorphic. If we would rather have the annihilators to be holomorphic, we need to change the complex structure. This can be done by swapping $\Omega$ and $\bar{\Omega}$, and $I$ should be modified accordingly (by a sign). This can be done in two ways:
\begin{enumerate}
\item Change the sign of $J$: this will keep the relations (\ref{sun}) intact but will change the orientation (\ref{metric}).
\item Take rather $I_i^{j} = g^{jk} J_{ki}$ instead of (\ref{I}): this will keep the orientation but add a sign in the relation (\ref{sun}).
\end{enumerate}
The first change is the choice made in \cite{Tsimpis:2016bbq} while the second was chosen in \cite{Koerber:2010bx}. All cases are made to be consistent with the relation (\ref{IOmega}), which is also merely a convention. Fortunately, everyone seems to agree on this one.

%
%

\bibliography{refs}

\begin{thebibliography}{10}

\bibitem{Ossa:1992vc}
Xenia~C. de~la Ossa and Fernando Quevedo.
\newblock {Duality symmetries from nonAbelian isometries in string theory}.
\newblock {\em Nucl.Phys.}, B403:377--394, 1993.

\bibitem{Fridling:1983ha}
B.E. Fridling and A.~Jevicki.
\newblock {Dual Representations and Ultraviolet Divergences in Nonlinear
  $\sigma$ Models}.
\newblock {\em Phys.Lett.}, B134:70, 1984.

\bibitem{Fradkin:1984ai}
E.S. Fradkin and Arkady~A. Tseytlin.
\newblock {Quantum Equivalence of Dual Field Theories}.
\newblock {\em Annals Phys.}, 162:31, 1985.

\bibitem{Sfetsos:2010uq}
Konstadinos Sfetsos and Daniel~C. Thompson.
\newblock {On non-abelian T-dual geometries with Ramond fluxes}.
\newblock {\em Nucl.Phys.}, B846:21--42, 2011.

\bibitem{Lozano:2011kb}
Yolanda Lozano, Eoin. O~Colgain, Konstadinos Sfetsos, and Daniel~C. Thompson.
\newblock {Non-abelian T-duality, Ramond Fields and Coset Geometries}.
\newblock {\em JHEP}, 1106:106, 2011.

\bibitem{Itsios:2013wd}
Georgios Itsios, Carlos Nunez, Konstadinos Sfetsos, and Daniel~C. Thompson.
\newblock {Non-Abelian T-duality and the AdS/CFT correspondence:new N=1
  backgrounds}.
\newblock {\em Nucl.Phys.}, B873:1--64, 2013.

\bibitem{Lozano:2012au}
Yolanda Lozano, Eoin O~Colgain, Diego Rodriguez-Gomez, and Konstadinos Sfetsos.
\newblock {Supersymmetric $AdS_6$ via T Duality}.
\newblock {\em Phys.Rev.Lett.}, 110(23):231601, 2013.

\bibitem{Lozano:2014ata}
Yolanda Lozano and Niall~T. Macpherson.
\newblock {A new AdS$_{4}$/CFT$_{3}$ dual with extended SUSY and a spectral
  flow}.
\newblock {\em JHEP}, 11:115, 2014.

\bibitem{Bea:2015fja}
Yago Bea, Jose~D. Edelstein, Georgios Itsios, Karta~S. Kooner, Carlos Nunez,
  Daniel Schofield, and J.~Anibal Sierra-Garcia.
\newblock {Compactifications of the Klebanov-Witten CFT and new AdS$_{3}$
  backgrounds}.
\newblock {\em JHEP}, 05:062, 2015.

\bibitem{Lozano:2015bra}
Yolanda Lozano, Niall~T. Macpherson, Jes{\'u}s Montero, and Eoin~{\'O}
  Colg{\'a}in.
\newblock {New $AdS_3 \times S^2$ T-duals with $ \mathcal{N}=\left(0,4\right) $
  supersymmetry}.
\newblock {\em JHEP}, 08:121, 2015.

\bibitem{Macpherson:2014eza}
Niall~T. Macpherson, Carlos N\'u\~nez, Leopoldo~A. Pando~Zayas, Vincent G.~J.
  Rodgers, and Catherine~A. Whiting.
\newblock {Type IIB supergravity solutions with AdS$_{5}$ from Abelian and
  non-Abelian T dualities}.
\newblock {\em JHEP}, 1502:040, 2015.

\bibitem{Zayas:2015azn}
Leopoldo~A. Pando~Zayas, Vincent G.~J. Rodgers, and Catherine~A. Whiting.
\newblock {Supergravity solutions with AdS$_{4}$ from non-Abelian T-dualities}.
\newblock {\em JHEP}, 02:061, 2016.

\bibitem{PandoZayas:2017ier}
Leopoldo~A. Pando~Zayas, Dimitrios Tsimpis, and Catherine~A. Whiting.
\newblock {Supersymmetric IIB background with AdS$_4$ vacua from massive IIA
  supergravity}.
\newblock {\em Phys. Rev.}, D96(4):046013, 2017.

\bibitem{Lozano:2016wrs}
Yolanda Lozano, Niall~T. Macpherson, Jesus Montero, and Carlos Nunez.
\newblock {Three-dimensional $ \mathcal{N}=4 $ linear quivers and non-Abelian
  T-duals}.
\newblock {\em JHEP}, 11:133, 2016.

\bibitem{Haack:2009jg}
Michael Haack, Dieter Lust, Luca Martucci, and Alessandro Tomasiello.
\newblock {Domain walls from ten dimensions}.
\newblock {\em JHEP}, 10:089, 2009.

\bibitem{Lust:2009mb}
Dieter L\"{u}st and Dimitrios Tsimpis.
\newblock {New supersymmetric AdS(4) type II vacua}.
\newblock {\em JHEP}, 09:098, 2009.

\bibitem{Rota:2015aoa}
Andrea Rota and Alessandro Tomasiello.
\newblock {AdS${}_4$ compactifications of AdS${}_7$ solutions in type II
  supergravity}.
\newblock {\em arXiv:1502.06622}, 2015.

\bibitem{Youm:1997hw}
Donam Youm.
\newblock {Black holes and solitons in string theory}.
\newblock {\em Phys. Rept.}, 316:1--232, 1999.

\bibitem{Hanany:1996ie}
Amihay Hanany and Edward Witten.
\newblock {Type IIB superstrings, BPS monopoles, and three-dimensional gauge
  dynamics}.
\newblock {\em Nucl. Phys.}, B492:152--190, 1997.

\bibitem{Lozano:2016kum}
Yolanda Lozano and Carlos N{\'u}{\~n}ez.
\newblock {Field theory aspects of non-Abelian T-duality and $ \mathcal{N} =$ 2
  linear quivers}.
\newblock {\em JHEP}, 05:107, 2016.

\bibitem{Gaiotto:2009gz}
Davide Gaiotto and Juan Maldacena.
\newblock {The Gravity duals of N=2 superconformal field theories}.
\newblock {\em JHEP}, 10:189, 2012.

\bibitem{Itsios:2017cew}
Georgios Itsios, Yolanda Lozano, Jesus Montero, and Carlos Nunez.
\newblock {The AdS$_{5}$ non-Abelian T-dual of Klebanov-Witten as a $
  \mathcal{N}=1 $ linear quiver from M5-branes}.
\newblock {\em JHEP}, 09:038, 2017.

\bibitem{Lozano:2013oma}
Yolanda Lozano, Eoin {\'O}~Colg{\'a}in, and Diego Rodr{\'\i}guez-G{\'o}mez.
\newblock {Hints of 5d Fixed Point Theories from Non-Abelian T-duality}.
\newblock {\em JHEP}, 05:009, 2014.

\bibitem{Lozano:2015cra}
Yolanda Lozano, Niall~T. Macpherson, and Jes{\'u}s Montero.
\newblock {A $ \mathcal{N}=2 $ supersymmetric AdS$_{4}$ solution in M-theory
  with purely magnetic flux}.
\newblock {\em JHEP}, 10:004, 2015.

\bibitem{Itsios:2017nou}
Georgios Itsios, Horatiu Nastase, Carlos N{\'u}{\~n}ez, Konstantinos Sfetsos,
  and Salom{\'o}n Zacar{\'\i}as.
\newblock {Penrose limits of Abelian and non-Abelian T-duals of $AdS_5\times
  S^5$ and their field theory duals}.
\newblock {\em JHEP}, 01:071, 2018.

\bibitem{Gauntlett:2004hh}
Jerome~P. Gauntlett, Dario Martelli, James~F. Sparks, and Daniel Waldram.
\newblock {A New infinite class of Sasaki-Einstein manifolds}.
\newblock {\em Adv. Theor. Math. Phys.}, 8(6):987--1000, 2004.

\bibitem{Martelli:2004wu}
Dario Martelli and James Sparks.
\newblock {Toric geometry, Sasaki-Einstein manifolds and a new infinite class
  of AdS/CFT duals}.
\newblock {\em Commun. Math. Phys.}, 262:51--89, 2006.

\bibitem{Martelli:2008rt}
Dario Martelli and James Sparks.
\newblock {Notes on toric Sasaki-Einstein seven-manifolds and AdS(4)/CFT(3)}.
\newblock {\em JHEP}, 11:016, 2008.

\bibitem{Petrini:2009ur}
Michela Petrini and Alberto Zaffaroni.
\newblock {N=2 solutions of massive type IIA and their Chern-Simons duals}.
\newblock {\em JHEP}, 09:107, 2009.

\bibitem{Gaiotto:2009yz}
Davide Gaiotto and Alessandro Tomasiello.
\newblock {Perturbing gauge/gravity duals by a Romans mass}.
\newblock {\em J. Phys.}, A42:465205, 2009.

\bibitem{Passias:2018zlm}
Achilleas Passias, Dani{\"e}l Prins, and Alessandro Tomasiello.
\newblock {A massive class of $\mathcal{N} = 2$ AdS$_4$ IIA solutions}.
\newblock {\em JHEP}, 10:071, 2018.

\bibitem{Aharony:2010af}
Ofer Aharony, Daniel Jafferis, Alessandro Tomasiello, and Alberto Zaffaroni.
\newblock {Massive type IIA string theory cannot be strongly coupled}.
\newblock {\em JHEP}, 11:047, 2010.

\bibitem{Tomasiello:2010zz}
Alessandro Tomasiello and Alberto Zaffaroni.
\newblock {Parameter spaces of massive IIA solutions}.
\newblock {\em JHEP}, 04:067, 2011.

\bibitem{Guarino:2015jca}
Adolfo Guarino, Daniel~L. Jafferis, and Oscar Varela.
\newblock {String Theory Origin of Dyonic N=8 Supergravity and Its Chern-Simons
  Duals}.
\newblock {\em Phys. Rev. Lett.}, 115(9):091601, 2015.

\bibitem{Itsios:2012dc}
Georgios Itsios, Yolanda Lozano, Eoin. O~Colgain, and Konstadinos Sfetsos.
\newblock {Non-Abelian T-duality and consistent truncations in type-II
  supergravity}.
\newblock {\em JHEP}, 08:132, 2012.

\bibitem{Kelekci:2014ima}
\"{O}zg\"{u}r Kelekci, Yolanda Lozano, Niall~T. Macpherson, and Eoin~{\'O}
  Colg{\'a}in.
\newblock {Supersymmetry and non-Abelian T-duality in type II supergravity}.
\newblock {\em Class. Quant. Grav.}, 32(3):035014, 2015.

\bibitem{Macpherson:2017mvu}
Niall~T. Macpherson, Jes{\'u}s Montero, and Dani{\"e}l Prins.
\newblock {Mink $_3\times S^3$ solutions of type II supergravity}.
\newblock {\em Nucl. Phys.}, B933:185--233, 2018.

\bibitem{Koerber:2010bx}
Paul Koerber.
\newblock {Lectures on Generalized Complex Geometry for Physicists}.
\newblock {\em Fortsch. Phys.}, 59:169--242, 2011.

\bibitem{Tsimpis:2016bbq}
Dimitrios Tsimpis.
\newblock {Generalized geometry lectures on type II backgrounds}.
\newblock {\em arXiv:1606.08674}, 2016.

\end{thebibliography}
\bibliographystyle{unsrt}
\end{document}